\documentclass[a4paper,11pt]{article}
\pdfoutput=1 

\usepackage{jinstpub} 

\title{A fast and efficient SIMD track reconstruction algorithm for the LHCb Upgrade 1 VELO-PIX detector}

\author[1,2]{A. Hennequin}
\author[1]{B. Couturier}
\author[3]{V. V. Gligorov}
\author[1]{S. Ponce}
\author[3]{R. Quagliani}
\author[2]{L. Lacassagne}

\affiliation[1]{LHCb Experiment, CERN, Geneva, Switzerland}
\affiliation[2]{LIP6, Sorbonne Universit{\'e}, CNRS, Paris, France}
\affiliation[3]{LPNHE, Sorbonne Universit{\'e}, Paris Diderot Sorbonne Paris Cit{\'e}, CNRS/IN2P3, Paris, France}

\emailAdd{arthur.hennequin@cern.ch}

\abstract{
The upgraded CERN LHCb detector, due to start data taking in 2021, will have to reconstruct 4 TB/s of raw detector data in real time using commodity processors. This is one of the biggest real-time data processing challenges in any scientific domain.
We present an intrinsically parallel reconstruction algorithm for the vertex detector of the LHCb experiment designed to optimally exploit multi-core general purpose architectures. We compare it to previous state-of-the-art scalar pattern recognition algorithms and show significantly faster processing and in some cases increased physics performance over all current alternatives. We evaluate the algorithm on two high-end architectures from two different vendors and discuss in detail the impact of different SIMD Instruction Set Architecture extensions on the performance.
}


\begin{document}
\maketitle
\flushbottom

\section{Introduction}

The LHCb detector~\cite{Aaij:2014jba} is a general purpose spectrometer in the forward direction, optimized for the study of heavy flavour (primarily beauty and charm) hadrons at the Large Hadron Collider\footnote{The large hadron collider is an experimental facility located at CERN, which collides beams of protons with an energy of 7~TeV per beam and studies the products of those collisions. LHCb is one of the four main experiments located at the LHC.} (LHC). It also has a rich program of beyond Standard Model searches as well as hadronic, heavy ion, and electroweak physics. A key limitation of the current LHCb detector is that the full detector can only be read out at 1 MHz, while LHC collisions occur at 30~MHz. For this reason LHCb employs a hardware trigger system based on FPGAs, which has access to a limited subset of detector information at 30~MHz, to decide whether to read out a particular collision. The limited information available to this hardware trigger, and in particular the lack of information on charged particle trajectories (tracking) limits the efficiency for hadronic signals to roughly between 10\% and 30\%. \\

\begin{figure}[h]
\begin{center}
\includegraphics[width=\textwidth]{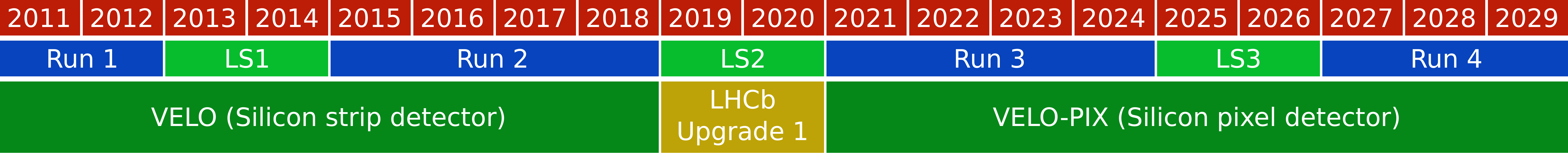}
\caption{LHC timeline and VELO detectors lifespan.}
\label{fig:timeline}
\end{center}
\end{figure}

For this reason, during the LHC second long shutdown (LS2), the LHCb detector will undergo its first upgrade, as shown in Figure~\ref{fig:timeline}. The maximal physics reach of collider experiments like LHCb is directly proportional to the number of proton collisions which the detector can record per second. The primary goal of this upgrade is to allow LHCb to increase the number of collisions per second by a factor five. In addition LHCb is switching to a full detector readout capable of operating at 30 MHz. The full 4 TB/s data rate will be sent for real-time processing in an off-the-shelf data center, where a two-stage software trigger\footnote{High Level Trigger, or HLT, in High Energy Physics jargon} will process the data in real time and select interesting signals for further offline physics analysis. The first High Level Trigger stage (HLT1) performs a fast reconstruction and selects approximately the 1~MHz of most interesting LHC collisions for further analysis. The second stage (HLT2), operating on the output of HLT1, performs a full reconstruction and real-time analysis of the data.  Removing the previous hardware trigger, which was reducing the data volume by a factor 20, implies that the overall data volume processed will increase by a factor of around 100. On the other hand if this data can be processed, the additional information available for early decision making will increase~\cite{Fitzpatrick:2244313} processing efficiency for key physics benchmarks by a factor between 2 and 10.\\

In order to fully exploit this additional information, it is necessary to improve LHCb's reconstruction algorithms so that they are able to process the LHCb upgrade data volume within the available computing resources. Concretely, the data rate seen by the upgraded LHCb HLT is around 100 times larger than for the current detector, with a flat budget for computing resources. In this paper, we present an efficient and highly parallel CPU implementation of one of the key LHCb reconstruction algorithms, we show that it fits within the available resources for the LHCb upgrade, and we discuss the scaling of its performance with some Single Instruction Multiple Data (SIMD) Instruction Set Architecture (ISA) extensions on Intel's Skylake and AMD's Zen2 architectures.\\

\subsection{The LHCb Upgrade 1 VELO-PIX detector}

\begin{figure}[h]
\begin{center}
\includegraphics[scale=0.4]{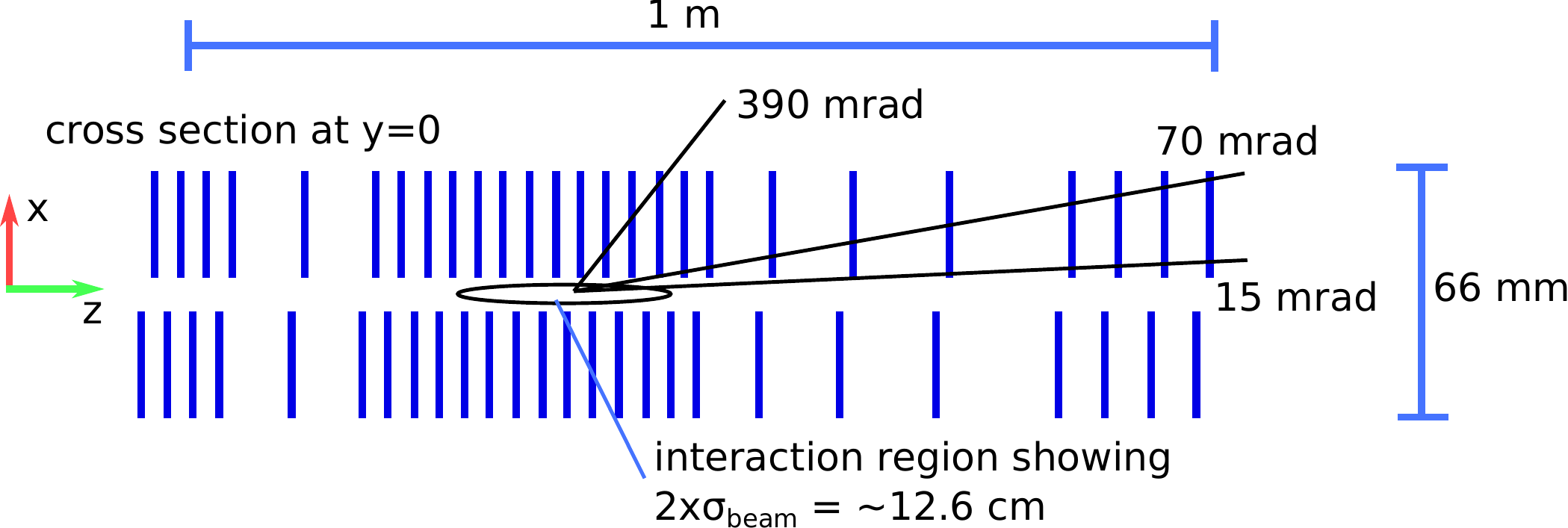}
\includegraphics[scale=0.4]{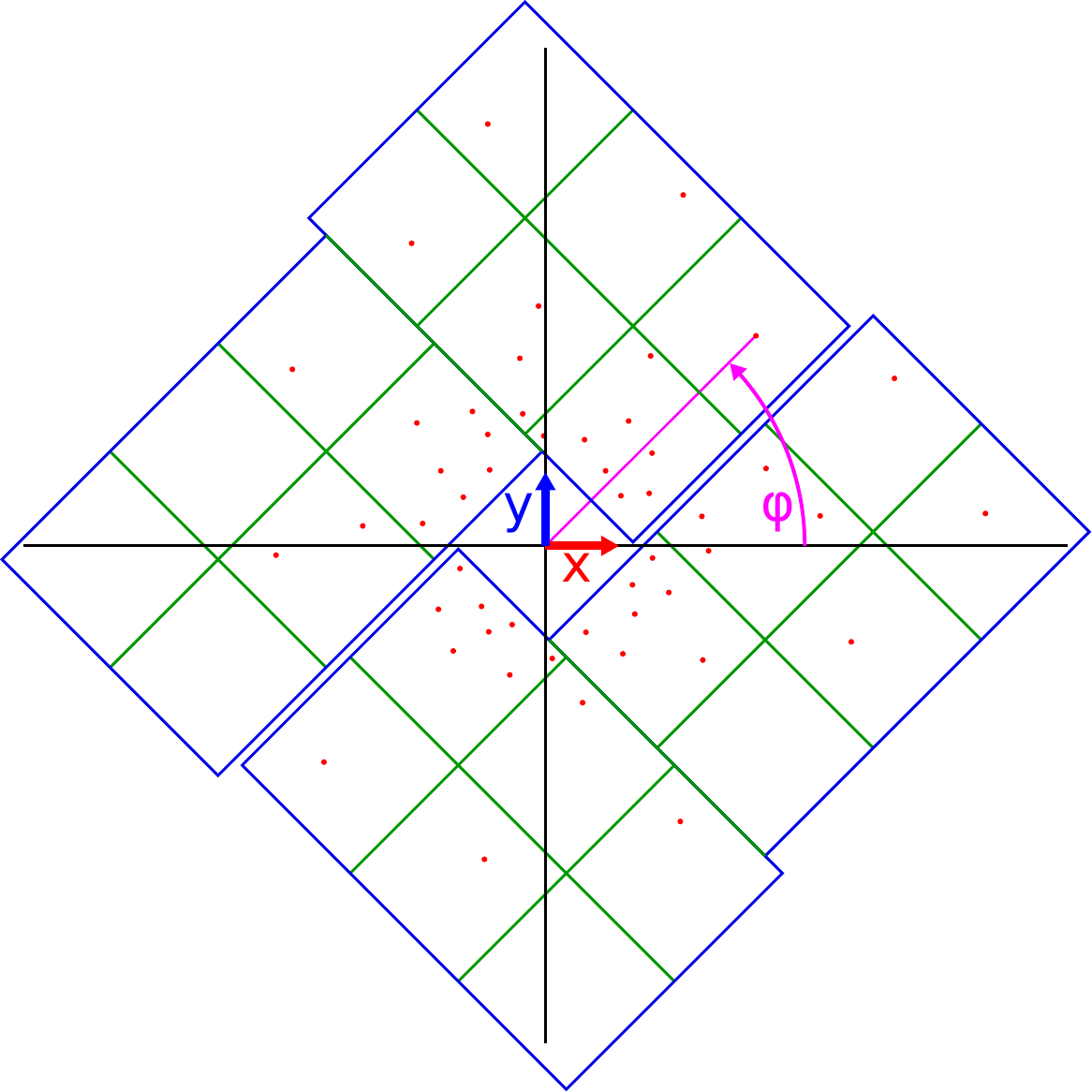}
\caption{VELO-PIX (Silicon pixel detector) Geometry. The (\textbf{x}, \textbf{y}, \textbf{z}) cartesian coordinate system is aligned with the beamline, with the \textbf{z} axis pointing downstream. $\phi$ is the angle around the beamline. In the right hand diagram, each square is a 14 $\times$ 14 mm silicon sensor of 256 $\times$ 256 pixels, positioned perpendicular to the beam.}
\label{fig:velopix_geom}
\end{center}
\end{figure}

In addition to the triggerless readout, one of the major changes in the upgraded LHCb detector is the total replacement of the VELO (Vertex Locator). The purpose of this detector, located around the beamline over the interaction region, is to precisely reconstruct the locations of the proton-proton collisions (primary vertices, PVs, in physics jargon) and separate tracks produced directly in PVs from tracks produced by particles which decay inside the VELO but away from the PV. In order to improve the resolution of the PV positions while increasing the luminosity, the detector technology changed from silicon strips to silicon pixels. The geometry of the new VELO-PIX detector is shown on Figure~\ref{fig:velopix_geom}. 

\subsection{Pattern recognition and fit}

Pattern recognition for tracks consists in finding hits consistent with a single particle traversing a detector. For the VELO detector, we search for almost straight lines coming from the beamline. To evaluate the quality of the tracking algorithm, we compare the reconstructed tracks to the set of reconstructible tracks from the ground truth given by the Monte-Carlo simulation~\cite{Sjostrand:2007gs,Agostinelli:2002hh,Allison:2006ve,LHCb-PROC-2010-056}. For the VELO, a particle is reconstructible if it leaves at least 3 hits in the detector. A correctly reconstructed track is one that has more than 70\% of its hits created by a single true particle. If more than one track candidate is matched to the same true particle, it is interpreted as a \textit{clone}. A track candidate that could not be matched to any particle is a \textit{fake}. We define the efficiency, the clone rate and the fake rate as follow:

\[\text{efficiency} = \frac{\lvert\{reconstructed\}\rvert}{\lvert\{reconstructible\}\rvert}\]
\[\text{clone rate} = \frac{\lvert\{clones\}\rvert}{\lvert\{clones\}\cup\{reconstructed\}\rvert}\] 
\[\text{fake rate} = \frac{\lvert\{fakes\}\rvert}{\lvert\{fakes\}\cup\{reconstructed\}\rvert}\]

A good tracking algorithm should have the highest efficiency and the lowest clone and fake rates possible. As the efficiency only accounts for the track being found but not the quality of the reconstructed tracks, we define the hit efficiency as the fraction of hits from a true particle included in the reconstructed track. This metric should be as high as possible and is a good indicator of the quality of the reconstructed tracks.\footnote{Not all track hits are equally important from a physics point of view. In particular, missing the first hit on the track worsens the resolution on the track's physics parameters far more than missing a hit in the middle of the track. Similarly, missing the last hit on the track worsens the resolution for extrapolating the track to the rest of the LHCb detector.}\\

Once the track candidates are found, a Kalman fit is performed on the hits in order to define the state closest to the beamline and the state at the end of the VELO (z = 770mm). A state consists in the slope of the track ($t_x$, $t_y$) and the coordinates at ($x_0$, $y_0$, $z=0$). Due to multiple scattering, it is expected that the two produced states are slightly different. We define the $\chi^2$ of the track as:

\[\chi^2 = \sum_{h=(x,y,z)}^{\{hits\}} (x_0+h_z t_x - h_x)^2 + (y_0+h_z t_y - h_y)^2\]

The $\chi^2$ can later be used to remove fake tracks or be included in the vertex $\chi^2$ computation.

\subsection{History of VELO Tracking algorithms}

\begin{figure}[h]
\begin{center}
\includegraphics[scale=0.45]{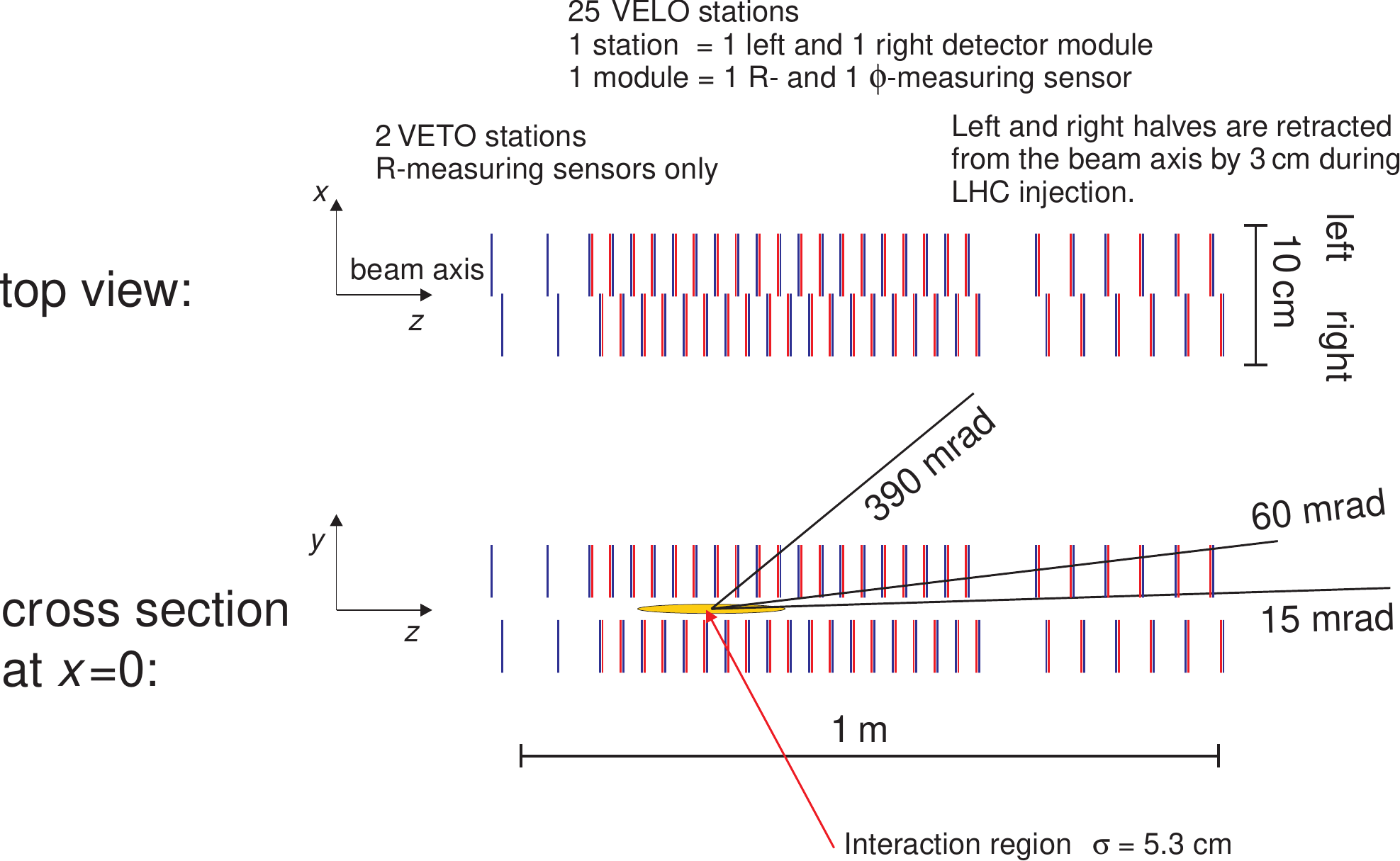}
\includegraphics[scale=0.3]{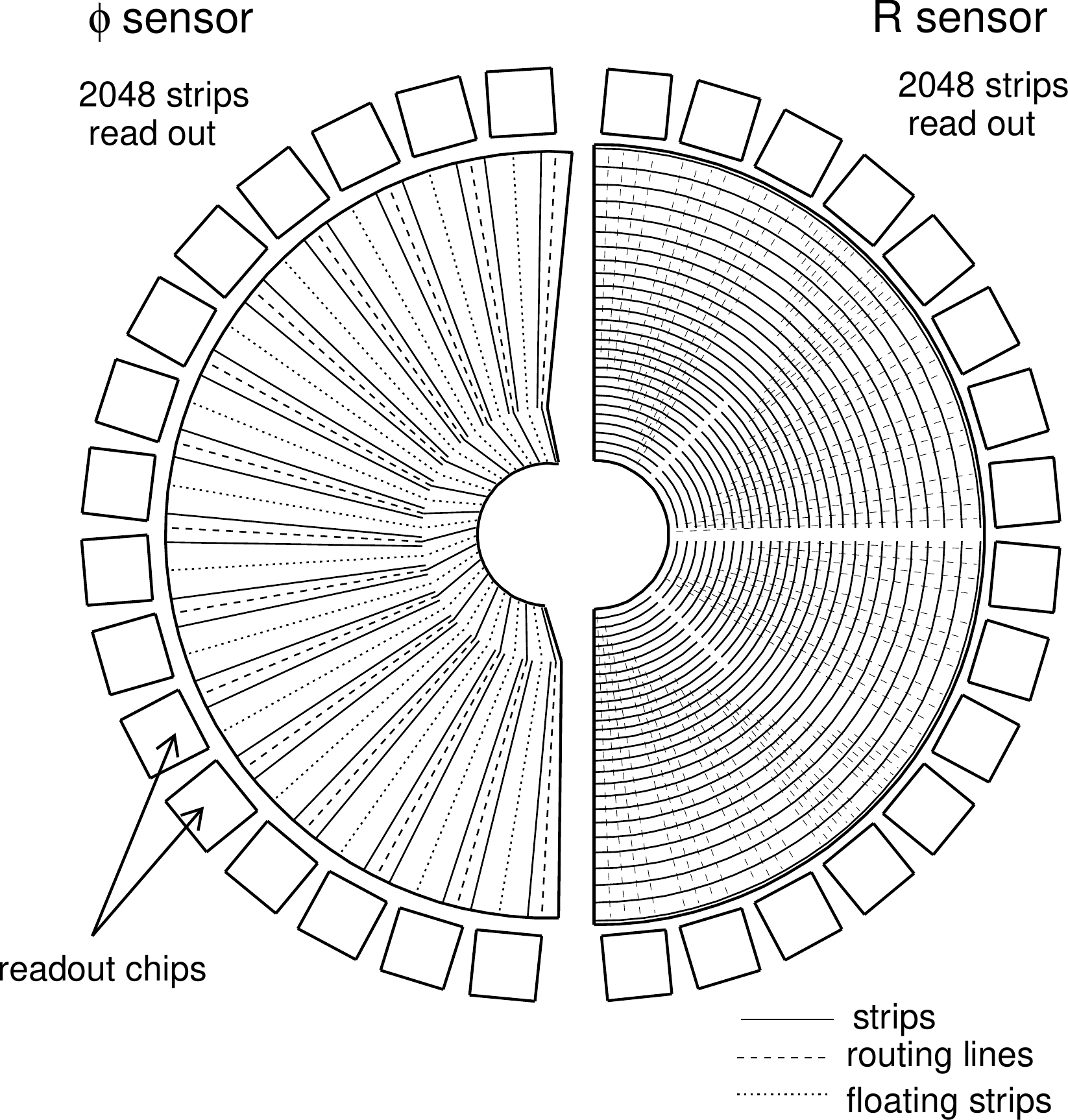}
\caption{VELO (Silicon strip detector) Geometry. Figures from \cite{LHCB-VELO-TDR2001}, the ``beam axis'' is what we refer to as beamline in the rest of the paper. The right hand diagram has an example of a $\phi$ sensor on the left and an \textbf{R} sensor on the right. The radius of the detectors is 45 mm.}
\label{fig:rphicartoon}
\end{center}
\end{figure}

The VELO Tracking algorithm used during the first LHC data taking period (Run 1) was developed in 2002 \cite{Callot2003}. At that time, the VELO detector geometry was using strips along $\phi$ and \textbf{R}, as shown in Figure~\ref{fig:rphicartoon}. Consequently, the tracking was done in two steps. First, a 2D tracking was performed in the \textbf{R-z} projection where it was easy to find interesting tracks based on the slope and alignments, then a space tracking step matched these \textbf{R-z} track stubs to clusters on the $\phi$ strips.\\

The 2D tracking began by finding a triplet of aligned clusters in three consecutive R sensors, then extended it as much as possible by predicting the radius in the next sensors and finding the closest cluster. To avoid finding the same track again, used clusters were marked and not considered in later searches. To avoid missing tracks due to sensor inefficiencies, the algorithm was allowed to skip one sensor in this search. The algorithm processed the R sensors in a single step going toward the interaction region. Due to the criteria applied on the slope and the single pass, no backward tracks were reconstructed. Subsequently, the space tracking was performed for every \textbf{R-z} track candidate. A second \textbf{R-z} tracking was then performed in the reverse direction to find the backward tracks, useful for finding primary vertices. After matching the \textbf{R-z} track stubs to clusters on the $\phi$ strips, the best track candidate was selected. The candidate with the highest total number of clusters was selected, or the candidate with the best track fit $\chi^2$ in case of equality.\\

In 2004, the algorithm was updated \cite{Callot2004} to fulfill the speed and efficiency requirements of the real-time reconstruction for the different LHCb trigger stages. The changes mainly concerned the tuning of tolerances and search windows. It was noted that large search windows were needed in the track extension step to allow the recovery of tracks not pointing to the beam-line, for which the \textbf{R-z} projection is not exactly a straight line, and low momentum tracks with large multiple scattering. In 2007, further tuning and analysis of the algorithm were performed \cite{Hutchcroft2007}.\\

In 2011, a new implementation of the algorithm was introduced \cite{Callot2011}, motivated by LHCb's choice to run at twice the design's instantaneous luminosity. The consequently higher number of proton-proton collisions per bunch crossing and detector occupancy, required the reconstruction to be reoptimized in order to fit into the constraints of LHCb's real-time data processing. In addition, during the 2010 run it was found that the VELO could not come as close to the LHC beamline as expected. Because of this, the search for \textbf{R-z} track stubs introduced a further inefficiency for tracks produced away from the beamline. The 2011 algorithm modified the \textbf{R-z} tracking to first search for quadruplets of clusters, then triplets of clusters among the remaining unused clusters. This approach allowed to reduce the fake track rate and sped the algorithm up. The rest of the algorithm remained very similar to the previous implementations. In 2015, a measurement of the reconstruction efficiency of the VELO tracking was published \cite{LHCb2015}.\\

 A new VELO-PIX (VP) detector has been developed for the LHCb upgrade~\cite{Williams2017}. This is a silicon detector based on planes of square pixels, with a broadly similar geometric coverage to the old \textbf{R-z} Velo. In 2009, the simulation framework started supporting the new detector and in 2012 the first version of the pixel VELO tracking algorithm was implemented \cite{Head2013}. In the pixel version, the input of the tracking algorithm are the 3D Cartesian coordinates of the reconstructed clusters on each pixel plane. Similar to the previous VELO Tracking algorithm, the tracks are seeded by looking for pairs of unused clusters whose estimated track slope would be compatible with the geometric acceptance of the other LHCb detector components. 
 Subsequently, the track candidates are extended upstream (smaller \textbf{z}-position) by extrapolating and looking for the closest cluster within a search window. A cut on the maximal scattering angle is added and the search is abandoned if no clusters are found on three consecutive stations. Three-clusters tracks are kept only if all their clusters are unused and their $\chi^2$ is below a parametrizable threshold. A detailed description of the algorithm and its performance was given in the VELO Upgrade 1 TDR \cite{LHCB-VELO-TDR2013}.\\
 
 A study was conducted to use vertical vectorization with 128-bit SSE SIMD extension to accelerate part of the algorithm \cite{Ticse2013} but resulted in a slowdown due, according to the authors, to the need for data preparation to take advantage of SIMD loads and stores. At that time, alternative global methods based on the Hough Transform \cite{Hough1959} and suitable for parallel architectures were also evaluated \cite{Ticse2013}\cite{Ristori:2000vg,Abba2014}. In 2014, a new local search algorithm based on triplet seeding on GPU was presented \cite{Badalov:1698101}. The main difference with the previous sequential work was that tracks were seeded and extended upstream independently, in parallel. A post-processing step cleaned ghosts. This preliminary work was further improved in \cite{Badalov2016CoprocessorIF}.\\

In 2018, the CPU pixel tracking algorithm was made faster in order to improve the throughput of the first stage of LHCb's real-time reconstruction \cite{DeCian2018}. The clusters were ordered by $\phi$ and the cluster search performed within $\phi$-windows, the search for backward and forward tracks was split in two different steps and some "speed-flags" were introduced, allowing early cuts in the track candidates. In 2019, the search by triplet algorithm was revisited for parallel architectures in the context of the Allen project \cite{CamporaPerez2019}. This new implementation used the $\phi$-windows to reduce the combinatorics and uses synchronization between each layer to avoid track overlap. 

\section{SIMD VELO Tracking}

\subsection{Data preparation}

\begin{figure}[h]
\begin{center}
\includegraphics[scale=0.5]{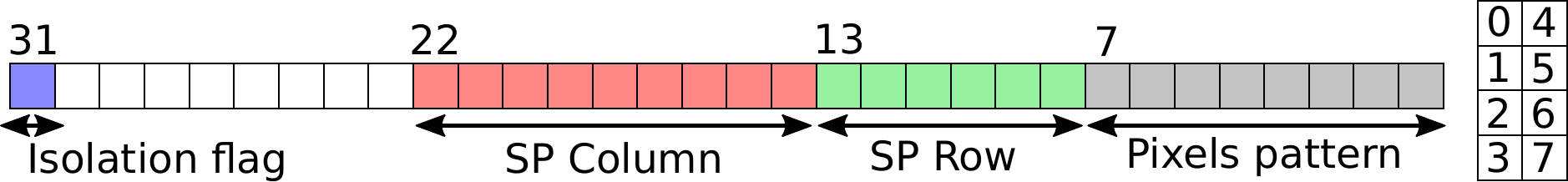}
\caption{VELO Super-pixel format}
\label{fig:sp_format}
\end{center}
\end{figure}

The VELO-PIX detector is divided into 52 L-shaped modules, as depicted in Figure~\ref{fig:velopix_geom}. Each module is itself composed by 4 sensors of 3 chips each. The chips have 256$\times$256 pixels, so the sensors have 256 rows and 768 columns. Each pixel is a square with a length of 55 microns (except at the chip border where the pixels are elongated). The sensor pixels are packed into Super-Pixels (SP) of size 2$\times$4 pixels, so the sensors have 64 SP rows and 384 SP columns~\cite{Hennessy2017}\cite{Poikela2015}. The modules are positioned along the z axis (see Figure~\ref{fig:velopix_geom}). Figure~\ref{fig:sp_format} shows the format of a Super-Pixel (SP) encoded in a 32-bit integer. The less significant byte is a bitmask representing the pixels. The row of the SP is stored from bit 8 to 13, and the column of the SP, from bit 14 to 22. The 31$^{th}$ bit is a flag indicating if the SP is isolated, ie. if it doesn't have any neighbours containing hits. The SP are delivered in small packet of bits called raw banks. There is one raw bank per sensor and each one contains the number of SP in the bank followed by the encoded SP. \\

\begin{figure}[h]
\begin{center}
\includegraphics[scale=0.5]{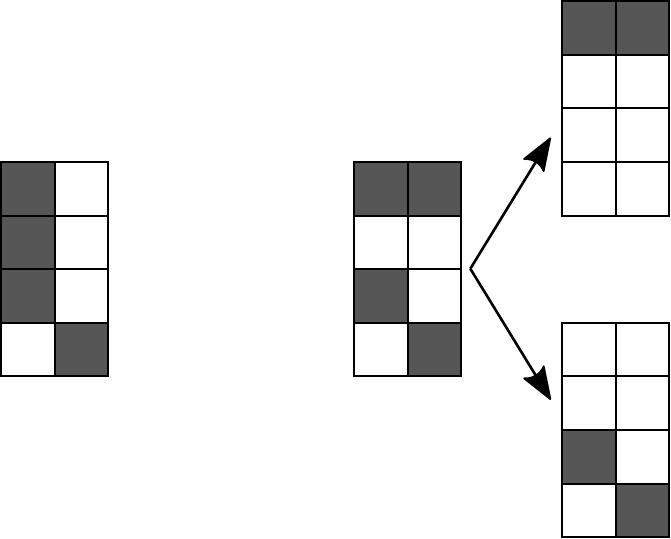}
\caption{A Super-Pixel containing one connected component (left) and a Super-Pixel containing two connected components, split in two SP (right)}
\label{fig:sp_patterns}
\end{center}
\end{figure}

Before doing any tracking, the raw pixels must be decoded and grouped into clusters. This step is crucial, as it prevents duplication of clusters if a particle hits the sensor at the boundary of multiple pixels. This clustering operation is a well known problem in computer vision, where it is referred as connected component labeling. Pioneer algorithms\cite{Rosenfeld1966,Haralick1981} have been accelerated continuously to match modern architectures like multi-core CPU, FPGA or GPU~\cite{He2017,Grana2018,Gupta2014,Cabaret2016,Hennequin2018_DASIP,Klaiber2016a}. But these algorithms were not adapted to low density images until they were specialized to take advantage of the data format~\cite{HennequinDASIP2019} to label a sparse list of SP instead of pixels. This allows to further reduce the amount of memory needed and to skip a decoding step. We first start by preparing the data: we remove the SPs that are known to be isolated and resolve them using lookup tables. For the remaining SPs, we test if there is more than one CC inside and split them if necessary. Figure~\ref{fig:sp_patterns} shows the two possible configurations for a SP: one connected component or two connected components. Because there can be at most two clusters per SP the maximum number of clusters in the image is $2\times$ the number of SPs. Once the SP list is prepared, we run the algorithm using a combination of bitwise operations and a lookup table to test the adjacency. Another lookup table is used for a fast computation of the first statistical moments and the number of pixels within a SP. Finally, the 2D clusters coordinates are transformed from their sensor space to the LHCb 3D global coordinate system, by multiplying them with the sensor's transformation matrix.

Each cluster is stored in a Structure of Array (SoA) layout of memory to ease the loading into SIMD registers. In addition to the 3D global position (x, y, z), we also store the approximated $\phi$ angle, computed as $\phi=atan2(y,x)$ using a fast atan2 function, and the LHCb identifier of the cluster.

\subsection{SIMD instructions}
\label{sec:SIMD}

\begin{figure}[!htb]
\centering
\includegraphics[scale=0.48]{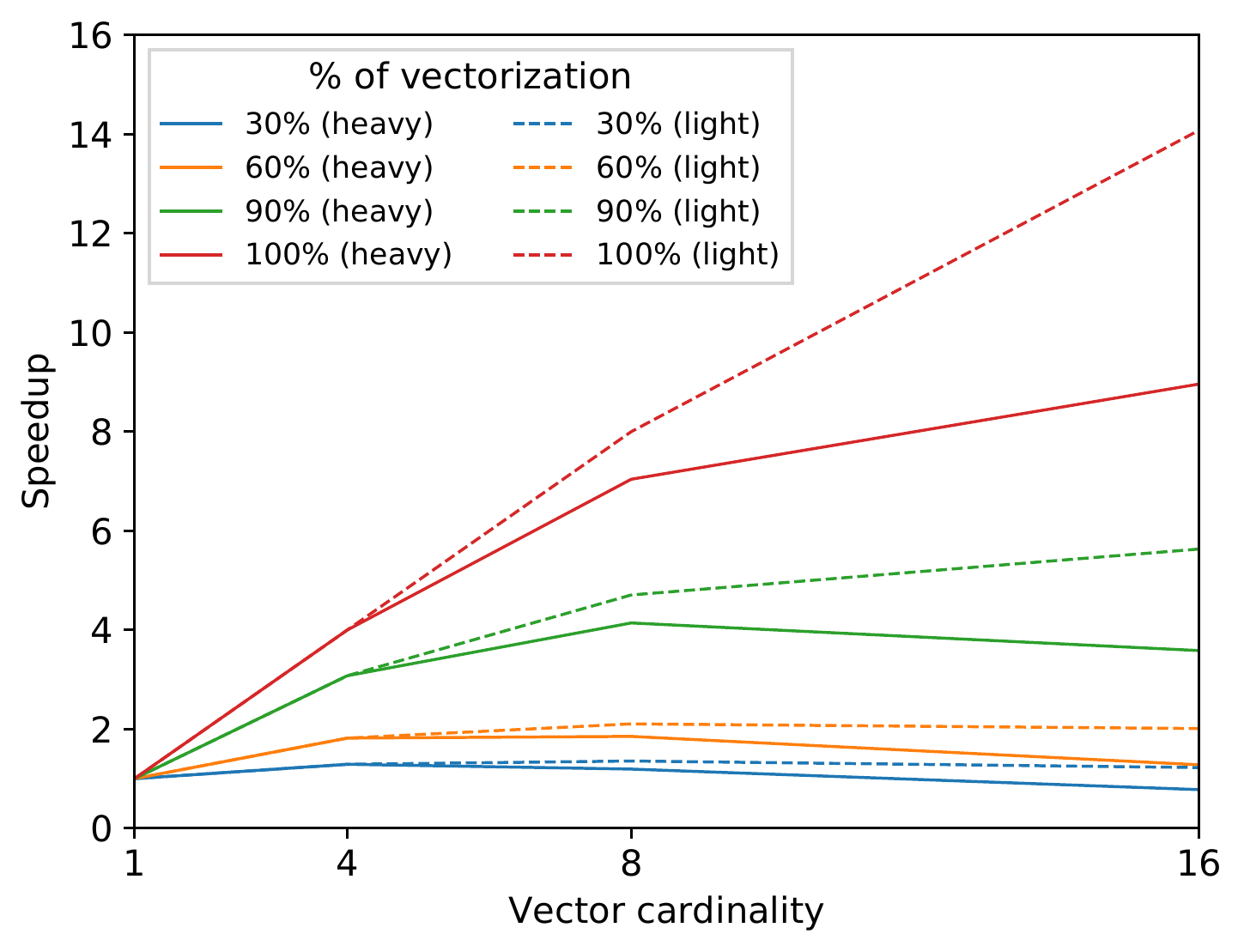}
\includegraphics[scale=0.48]{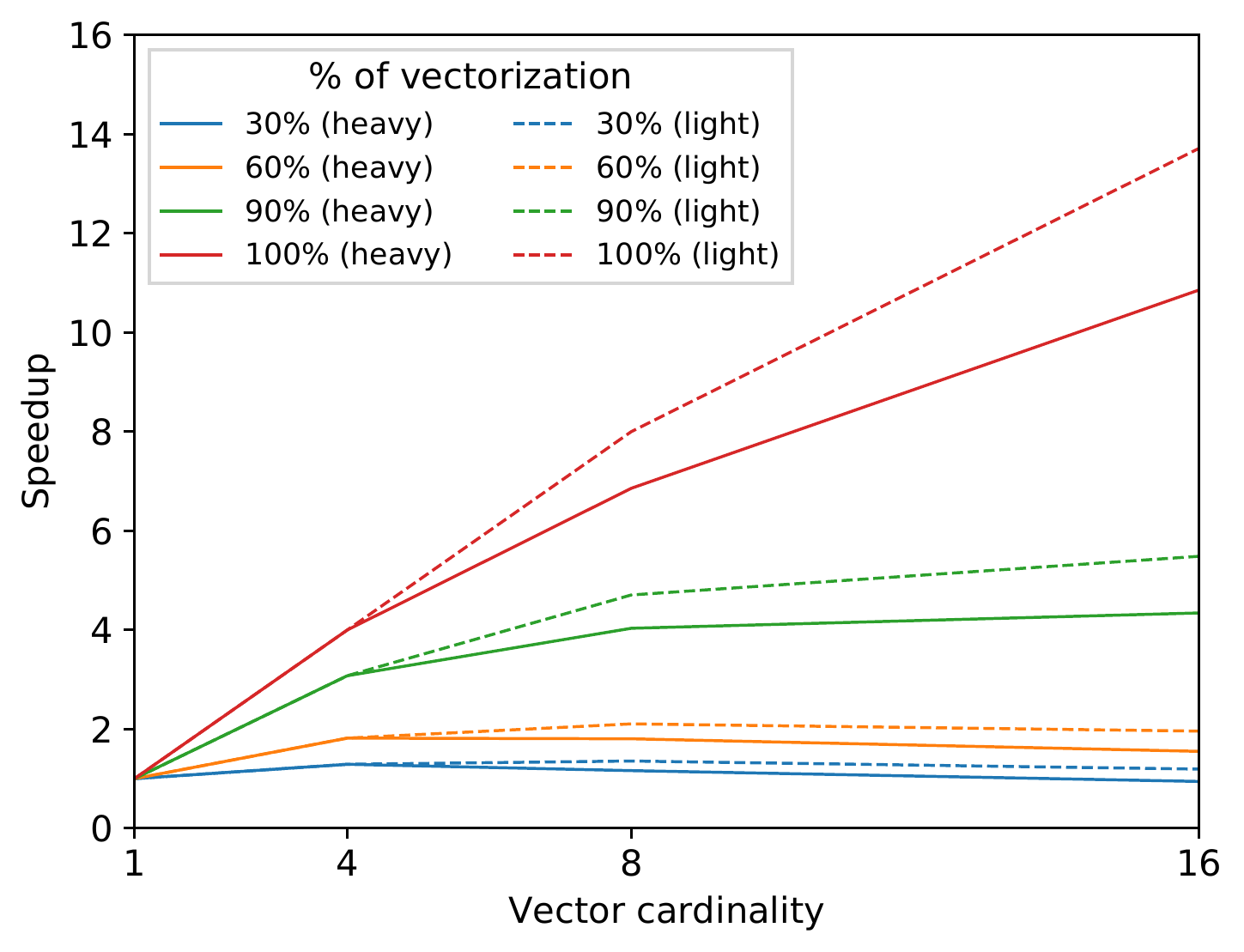}
\caption{Amdahl's law applied to SIMD vector width with frequency correction, for an Intel Xeon Silver 4114 (left) and an Intel Xeon Gold 6130 (right)}
\label{fig:amdahl_freq}
\end{figure}

As clock frequencies of modern processors are expected to stay near their current levels, or even to get lower, the primary method to improve the computation power of a chip is to increase either the number of processing units (cores) or the intrinsic parallelism of a core (SIMD). The speedup that can be achieved for a particular application depends on the amount of code that can be vectorized. Amdahl's law~\cite{Amdahl1967} gives a theoretical bound for the speedup:\\
\[speedup(c) = \frac{1}{1 - \tau + \frac{\tau}{c}}\]
where $c$ is the vector cardinality, and $\tau$ is the fraction of vectorized code. To reduce power consumption and help thermal stability, Intel CPUs use dynamic frequency scaling to limit the frequency of cores running SIMD instructions. There are three levels of frequency as shown in table~\ref{table:freq}. The frequency is reduced, per core, if the process encounter a sufficiently high density of instruction of the corresponding type. The frequency reduction consist in multiple steps. When the CPU detect that no heavy instructions are used anymore, it waits approximately 2~ms before reverting the changes: in the mean time, scalar code runs at the lowered frequency~\cite{Gottschlag2019}. If the application interleave scalar and AVX code, it will likely run at the AVX-induced lower frequency. We can modify Amdahl's law to account for this frequency scaling:
\[speedup(c) = \frac{1}{1 - \tau + \frac{\tau}{c}} \times \frac{freq(c)}{freq(1)}\]
$freq(c)$ is the maximum frequency for a vector of cardinality $c$. Figure~\ref{fig:amdahl_freq} shows the theoretical speedups with frequency correction, for light and heavy instructions, for two different Intel CPUs. As we can see, for wide vectors a large amount of vectorized code is needed to be able to keep increasing the performances. To counterbalance the effects of frequency scaling, vendors added more specific instructions that can carry out complex operations in fewer cycles.\\

\begin{table}[!htb]
\centering
\begin{tabular}{lll}
\hline
                           & Base & Turbo \\ \hline
Non-AVX / Light AVX2       & 2.1  & 2.8              \\
Heavy AVX2 / Light AVX-512 & 1.7  & 2.4              \\
Heavy AVX-512              & 1.3  & 1.9              \\ \hline
\end{tabular}
\caption{Base and maximum frequencies (GHz) for an Intel Xeon Gold 6130~\cite{xeon6130freq} for all cores active}
\label{table:freq}
\end{table}

\begin{figure}[h]
\centering
\includegraphics[scale=0.35]{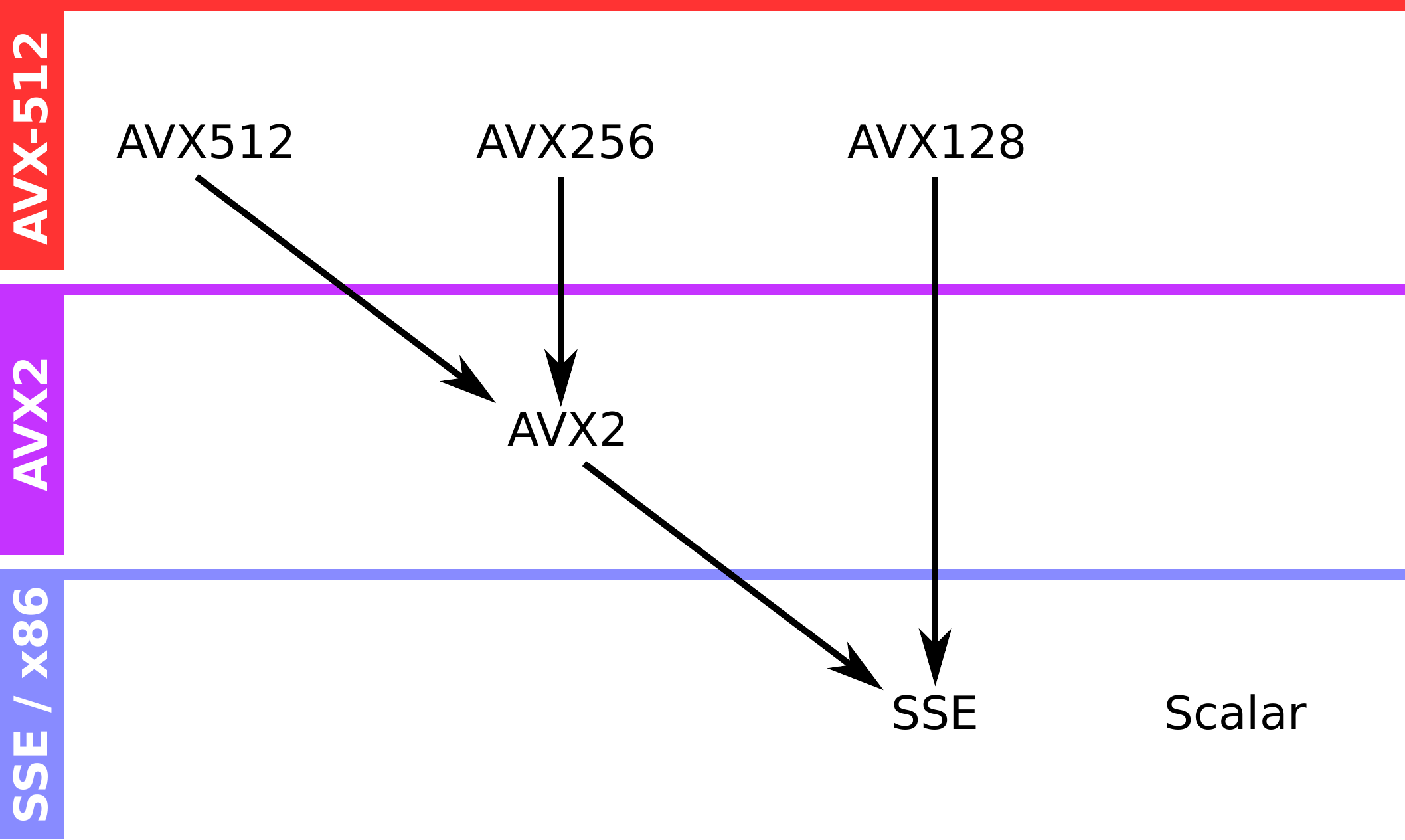}
\caption{SIMD wrappers fallback scheme on x86 architectures}
\label{fig:fallback}
\end{figure}

Developing new algorithms that efficiently use such a kind of SIMD architecture requires heavy algorithmic modifications and can only be done efficiently if the specific instructions are available. Due to this complex process, it is hard for a compiler to provide vectorization support for irregular algorithms. Domain Specific Languages (DSLs) such as Halide~\cite{ragan2013halide} or SPMD~\cite{Pharr2012} do not contain all patterns necessary for our problem and would also introduce significant additional complexity in the context of a physics codebase almost entirely written in C++. A less invasive option is to use SIMD libraries, like VC~\cite{Kretz2012}, UME::SIMD~\cite{Karpinski2017} or VecCore~\cite{Amadio2018}, that wrap the compiler intrinsics to provide a higher abstraction level to the developer. 
While these libraries work well for implementing most algorithms, they do not currently fully implement the latest SIMD ISA extensions. We therefore developed a set of vector length agnostic C++ template codes that can be instantiated for different SIMD backends in order to evaluate the impact of SIMD width on the performance and allow our algorithm to be ported on a wide range of architectures. In order to evaluate the impact of vector size on AVX-induced frequency scaling, we implemented an AVX256 and AVX128 backends that use AVX-512 instruction variants for 256 and 128 bits wide vector registers. This approach is similar to~\cite{Hennequin2019_WPMVP}. For simplicity, and because it matched this use case, the implementation is limited to 32-bit elements for integer and floating point types. The SIMD backend is determined by the developer, and is resolved at compile time following the fallback scheme depicted in Figure~\ref{fig:fallback}. This allows us to mix SIMD backends and provide easy debugging and testing capabilities while ensuring portability.

\subsection{Tracking algorithm} 

\begin{figure}[h]
\centering
\includegraphics[scale=0.25]{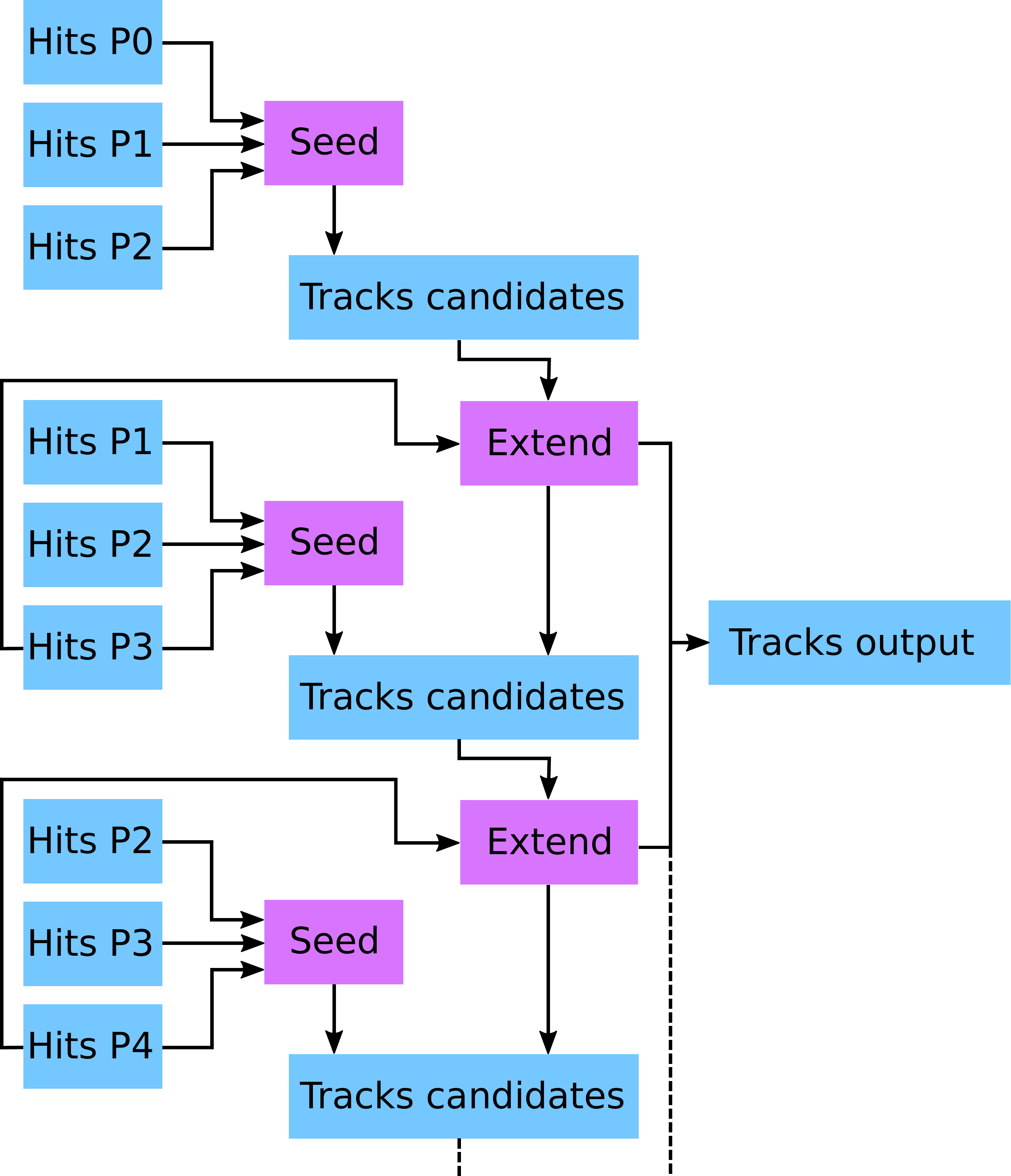}
\caption{Data flow within the algorithm. The hits are taken from the input planes P0 to P26, three by three, and processed in the seeding step to produce tracks candidates, which are then extended with the hits from the next layer. The candidates that could not have been extended are then copied in the tracks output container.}
\label{fig:velo_tracking}
\end{figure}

Like previous tracking algorithms, the proposed algorithm is a local search approach based on track following. To take advantage of track parallelism with SIMD, the algorithm is structured like the search by triplet algorithm \cite{CamporaPerez2019}, consisting in a track seeding and a track extending. Because on CPU the synchronization between vector elements is implicit, no synchronisation between layer is needed. Each layer's clusters are prepared on demand and stored in a small container with SoA layout. Three layers are needed at a time, so only three containers of clusters are allocated on the stack and we use pointer rotation to recycle them while moving through all VELO layers. This allows to reduce the algorithm memory footprint and to be more cache-friendly by improving data locality. Two track containers are used to memorize the track candidates created by the seeding and the finalized tracks produced by the track extending. Figure~\ref{fig:velo_tracking} shows the data flow within the algorithm.\\

\begin{figure}[h]
\centering
\includegraphics[scale=0.35]{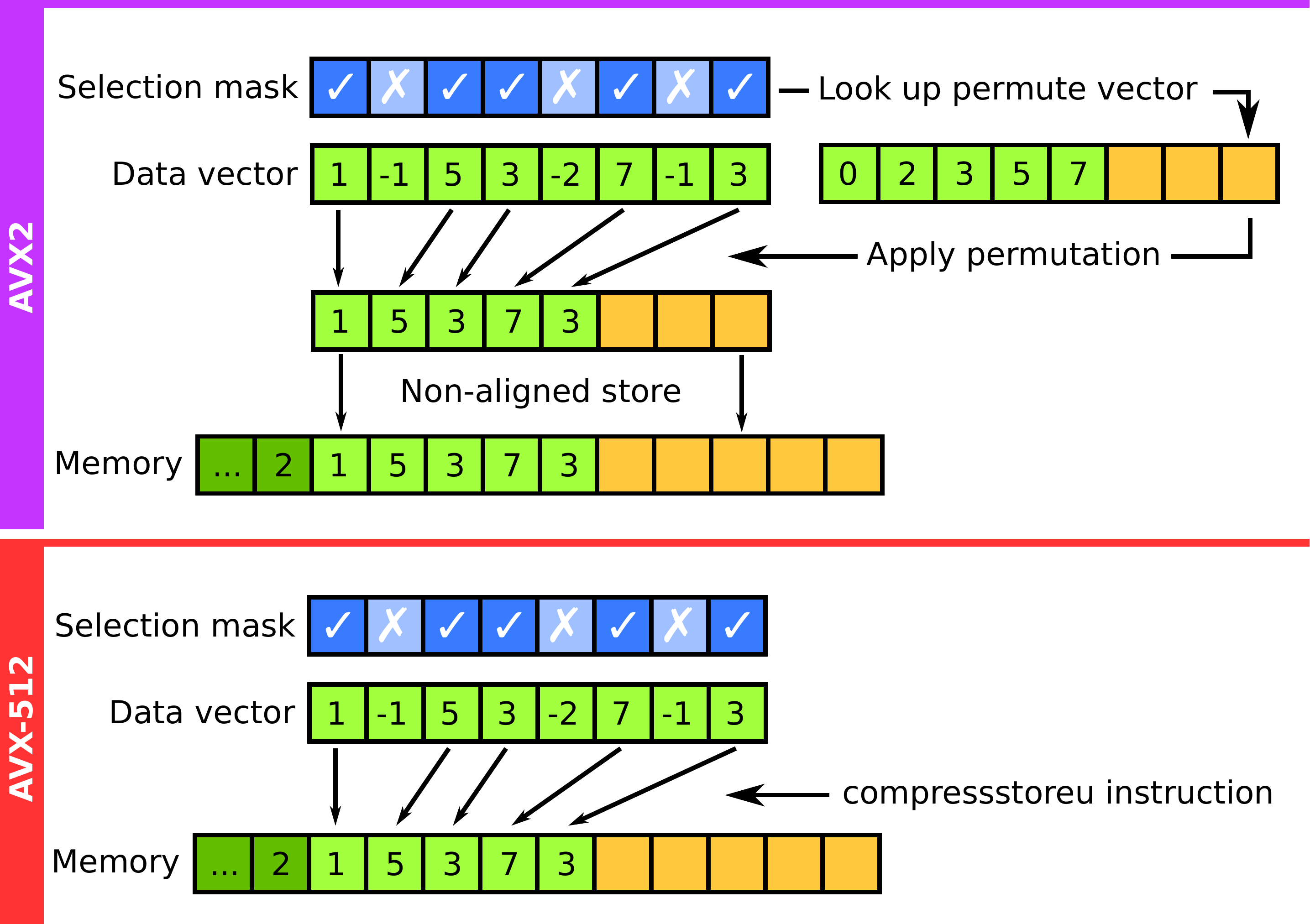}
\caption{Emulation of AVX-512's  \texttt{compressstoreu} instruction on AVX2 capable architecture.}
\label{fig:compress}
\end{figure}

One major algorithmic difference inspired by the SIMD requirements, is to compress data within each container of clusters and tracks. This ensures that useful data is always contiguous in memory, so that when loaded in SIMD registers no vector element is wasted. It also reduces the number of elements to process in the next steps. Instead of marking clusters like in previous algorithms using a boolean array, they are simply removed them from their container. Similarly, the tracks get moved from the candidate tracks container to the finalized tracks container or are discarded at the end of each extending step. This copy can be made efficiently using the \texttt{compressstoreu} instruction available in the AVX-512 instruction set. This instruction allows to pack some elements of a register to its left, based on a given mask and store it in memory. Figure~\ref{fig:compress} shows an example of a compression operation on AVX-512 and its emulation on AVX2 hardware. As this instruction is not available in older instruction sets, it can be emulated using a lookup-table and a permute instruction followed by a regular store~\cite{simdprune2019}. Because this instruction is used intensively, usually multiple times in a row with the same mask, the compiler is able to optimize the code to perform the lookup once and the core can pipeline the independent permute and store instructions, making it efficient.\\

\subsection{Seeding tracks}

\begin{figure}[h]
\centering
\includegraphics[scale=0.48]{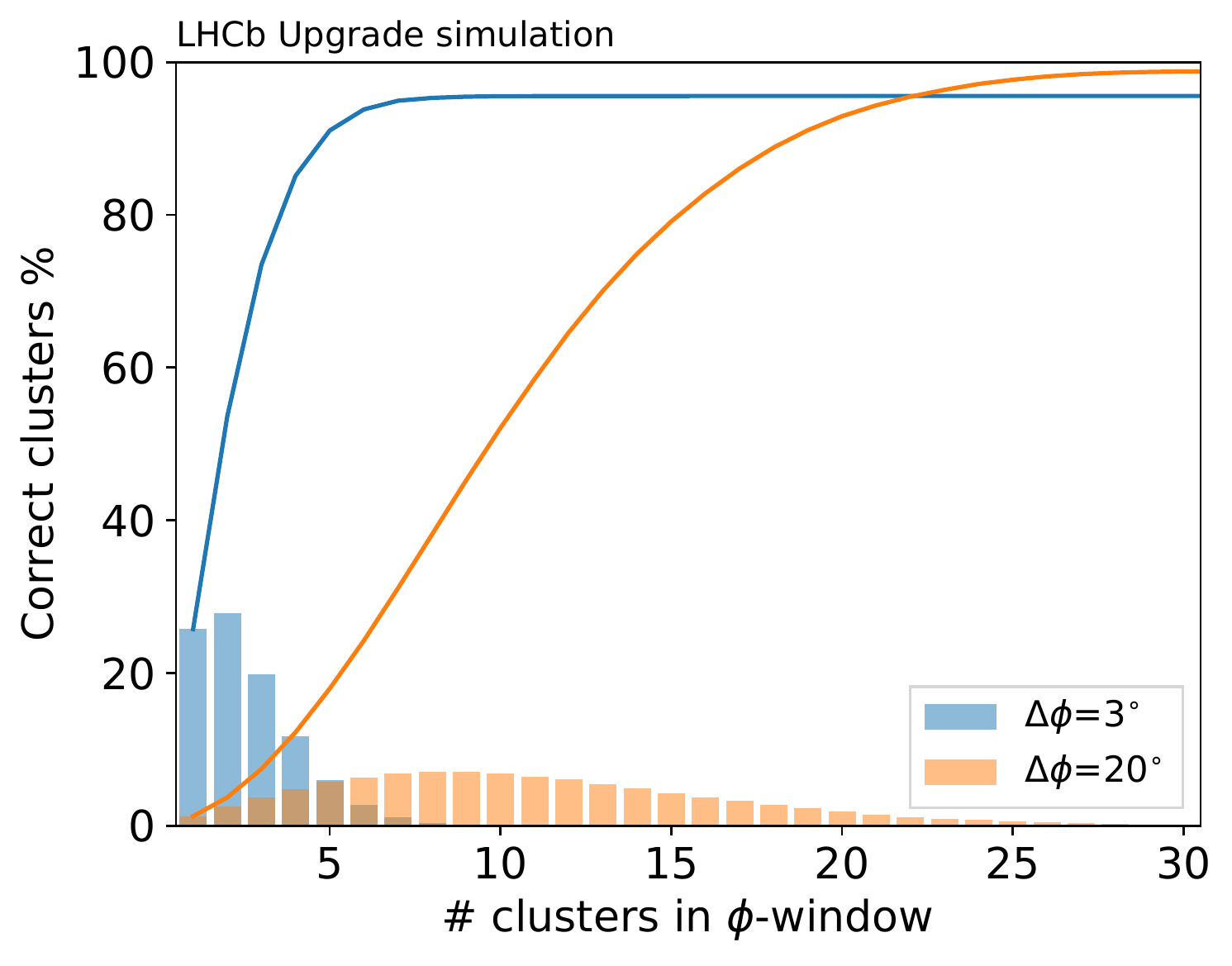}
\includegraphics[scale=0.48]{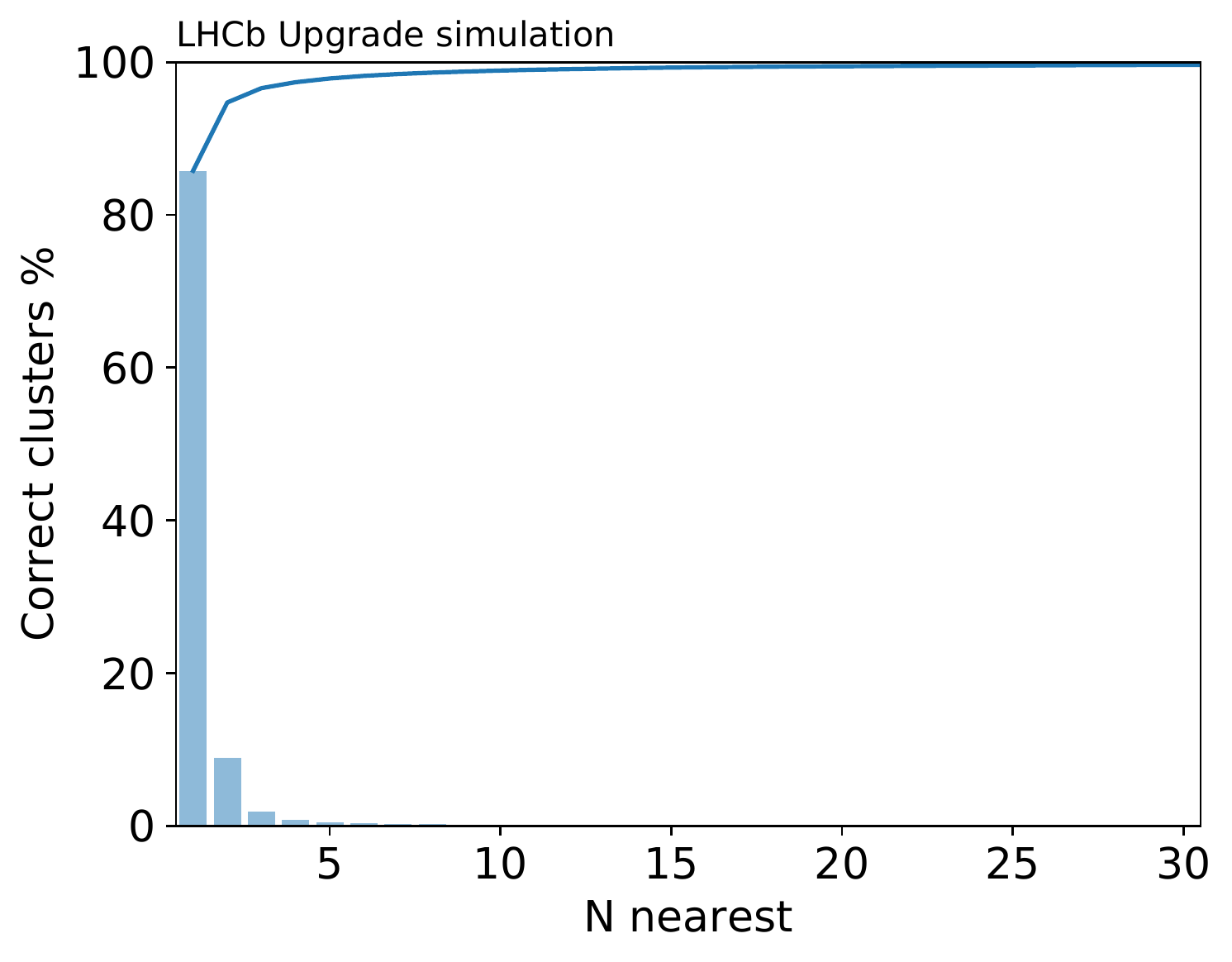}
\caption{On the left, the \% of correctly matched clusters for the $\phi$-window algorithms, depending on the number of cluster candidates ($3^{\circ}$ window is used in the "Fast" configuration of the Search by Pair algorithm and $20^{\circ}$ window is used in the "Best" configuration). On the right, the \% of correctly matched clusters for different number of candidates from our SIMD algorithm. These statistics are averaged on 100 Monte-Carlo simulated events, considering only the track seeding part of the algorithms and without marking used clusters.}
\label{fig:histo_closest}
\end{figure}

The track seeding is the most compute intensive part of the algorithm. In a typical upgrade event, one Velo layer contains an average of $\sim$100 clusters. Testing all possible triplet combination would require $O(100^3)$ tests. Previous algorithms based on pair and triplet search reduced the combinatorics by only testing pairs of clusters that have similar features. State-of-the-art algorithms build pair candidates by finding all second hits within a $\phi$-window around the first hit. This approach works well for the majority of tracks that comes from the beam line, but needs large tolerances to accept displaced tracks. Because a lot of LHCb's physics programs are based on these displaced tracks it is important to boost their reconstruction efficiency, even if they represent only a tiny fraction of reconstructible tracks. As shown on the left part of Figure~\ref{fig:histo_closest}, a $\phi$-window of $\pm3^{\circ}$ allows to correctly match 95.56\% of clusters by looking at a maximum of 10 clusters (2.6 in average), while a $\phi$-window of $\pm20^{\circ}$ is able to build 98.77\% of reconstructible pairs at the cost of having to test up to 30 candidates (10.8 in average). Apart from increasing the combinatorics, having a large candidate count increase the probability of generating fake tracks, and for these reasons we want to keep it as low as possible. Also, while having a variable number of candidates can be advantageous on a sequential architecture, a parallel architecture has to synchronize between processing elements so the time for all elements to finish is always the maximum of all elements' time.\\

Instead of using a $\phi$-window, we implemented a nearest in $\phi$ approach where we pick a fixed number of candidates N. As shown on the right part of Figure~\ref{fig:histo_closest}, we can match 96.59\% of clusters with only 3 candidates or 98.89\% of clusters with 10 candidates, reducing by a factor 3 the maximum number of candidates processed. This allows the SIMD algorithm to be more regular and less data dependent leading to a better utilisation of SIMD processing units. As the number of candidates N is small, they can fit in registers, avoiding costly memory accesses. The candidate positions and indices are stored in an N-sized array of SIMD register types.  As the number of candidates N is known at compile time, the compiler is able to fully unroll the loops (loop unwinding) over the candidates and the array to registers. If N is too large to fit all candidates in registers, the compiler generates spill code when it starts to overflow. By excluding used clusters when looking for the N nearest candidates, is allows more distant clusters to be tested if the closest ones are already used by another track, this helps to reconstruct displaced tracks in very dense layers. The limitation of processing multiple tracks at a time is that is allows the tracks to share some hits, potentially leading to clones. If the number of track processed in parallel is small, it doesn't have a big impact on the tracking efficiencies and clone rate. Because the hit container is not ordered the probability of sharing a hit among tracks processed in the same SIMD register is decreased. For larger SIMD registers widths, the conflict detection instructions available in AVX-512 can be used to remove the clones before propagating them. However, as the clone rate was already low, it was not necessary.\\

Once the initial pair candidates are built, they are extended in the third layer to search for the cluster minimizing the L2 distance to the linearly extrapolated position. In this step, all the hits are tested, without testing for the $\phi$ distance. As in previous algorithms, this L2-distance is called the \emph{scattering parameter} and the triplet candidate, built from the pair and the best cluster, that minimizes the scattering is kept. The triplet is accepted and added to the track candidate container if its scattering is lower than a configurable threshold value called \texttt{max\_scatter\_seeding}. When a cluster is used in a track, it is removed from the layer's cluster container.

\subsection{Extending tracks}

\begin{figure}[h]
\begin{center}
\includegraphics[scale=0.8]{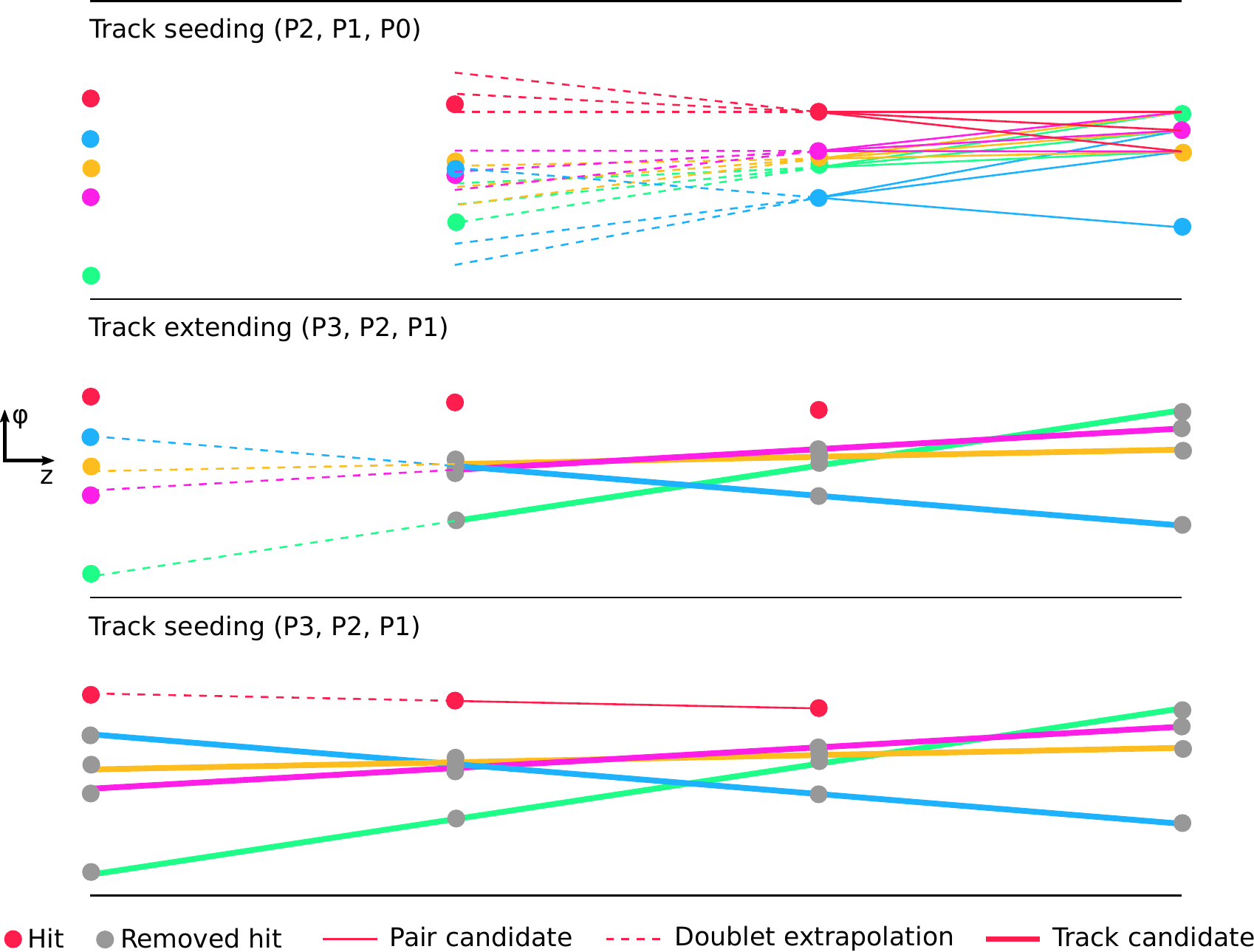}
\caption{The first seeding considers every hit in P1, builds 3 pair candidates with the nearest P0 hits in $\phi$ and extrapolates the doublet in P2 to find the P2 hit that minimizes the scattering. The used hits are removed, then every track candidate is extrapolated in P3 and extended if a hit is found. The first cycle is complete, and the algorithm performs another seeding in (P3,P2,P1), before continuing.}
\label{fig:track_example}
\end{center}
\end{figure}

After the seeding, the second step of an iteration of the tracking algorithm consist in finding hits in the current layer to extend the track candidates. Track candidates are processed in parallel and all non-used hits are tested using the same scattering criteria as in the seeding. The best hit is kept if the scattering is less than the \texttt{max\_scatter\_extending} threshold. To reduce the number of clones due to missing hit in layer, the track candidates not matched with a hit are kept one more iteration. The skipped layer counter is reset each time a hit is found. If the counter is exceeded once for a track candidate, it is moved to the output tracks container if it contains more than three hits or if the sum of the scatterings is less than a tighter threshold \texttt{max\_scatter\_3hits}, otherwise it is discarded. The partitioning of track candidates into next track candidates or track output is once again done using the compress-store pattern. Figure~\ref{fig:track_example} shows an example of track seeding and extension.

\section{Benchmarks}

\subsection{Benchmark procedure}

Following the approach of~\cite{DeCian2018}, all the algorithm configurations have been tested within the \texttt{GAUDI} framework~\cite{LHCB-COMPUTING-TDR2018}. The framework provides an efficient way of reading data from a local ramdisk and dispatching the events to the different threads. To decouple the single-threaded file I/O from the multi-threaded reconstruction, a double buffering technique was used where the first thread to reach the end of the current event buffer, swap it with the second one and refills the first from the next file. The sequential work of prefetching events consist in computing the pointer to the start of every event by doing the prefix sum of the event sizes. The decoding of the individual events raw banks is then performed by the thread in charge of the event reconstruction. The throughput is measured in number of events per second or, as it is analog to a frequency, in Hertz. To measure the time, the timing counter provided by the framework is used. It ignores the first and last 10\% of events for stability. All tested software was compiled with GCC 8.2. Unless explicitly specified, all results are given for a number of seeding pair candidate N=3.\\

Two setups were evaluated: a dual-socket Intel Xeon Gold 6130 and a single socket AMD EPYC "Rome" 7702. The Intel system features AVX-512, AVX2 and SSE instruction sets and scales its frequency according to Table~\ref{table:freq}. The AMD system only has AVX2 and SSE, at a frequency of 2.0 GHz.\footnote{All AMD throughput numbers in this article are given for 2.0 GHz.} The dual-socket Intel system has a total number of 32 physical cores and 64 threads, while the single socket AMD system has 64 physical cores and 128 threads. As every event is independent from the others, it is important to avoid memory latency induced by NUMA effects. On the Intel system we achieved optimal performance by launching one process per NUMA domain using the \texttt{numactl} utility. While the one socket AMD only has one NUMA domain for the whole chip, we found that best performance is achieved when launching one process per physical compute die, and thus always ran 8 independent jobs on this system.\\

In all tests, we run a full VELO reconstruction consisting in fetching the raw banks, applying a Global Event Cut of the 7\% biggest events, preparing the data, performing the actual tracking and fitting the resulting tracks. For the sake of fair comparison, the 2018 search by pair algorithm has been updated to the same plain old data input/output as the new algorithm, resulting in a speed improvement of $\sim$40\%.

\subsection{Throughput}

\begin{figure}[h!]
\begin{center}
\includegraphics[scale=0.48]{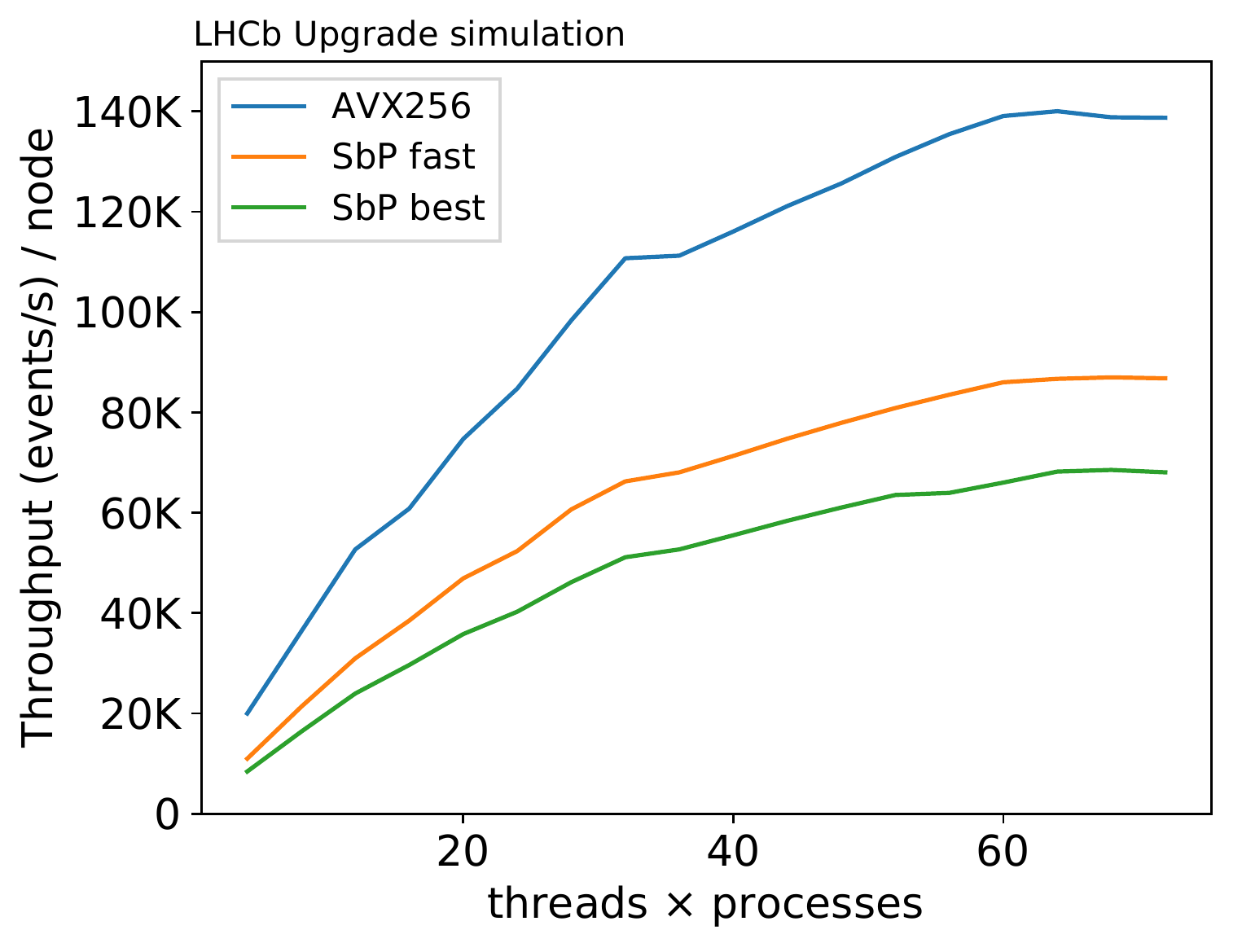}
\includegraphics[scale=0.48]{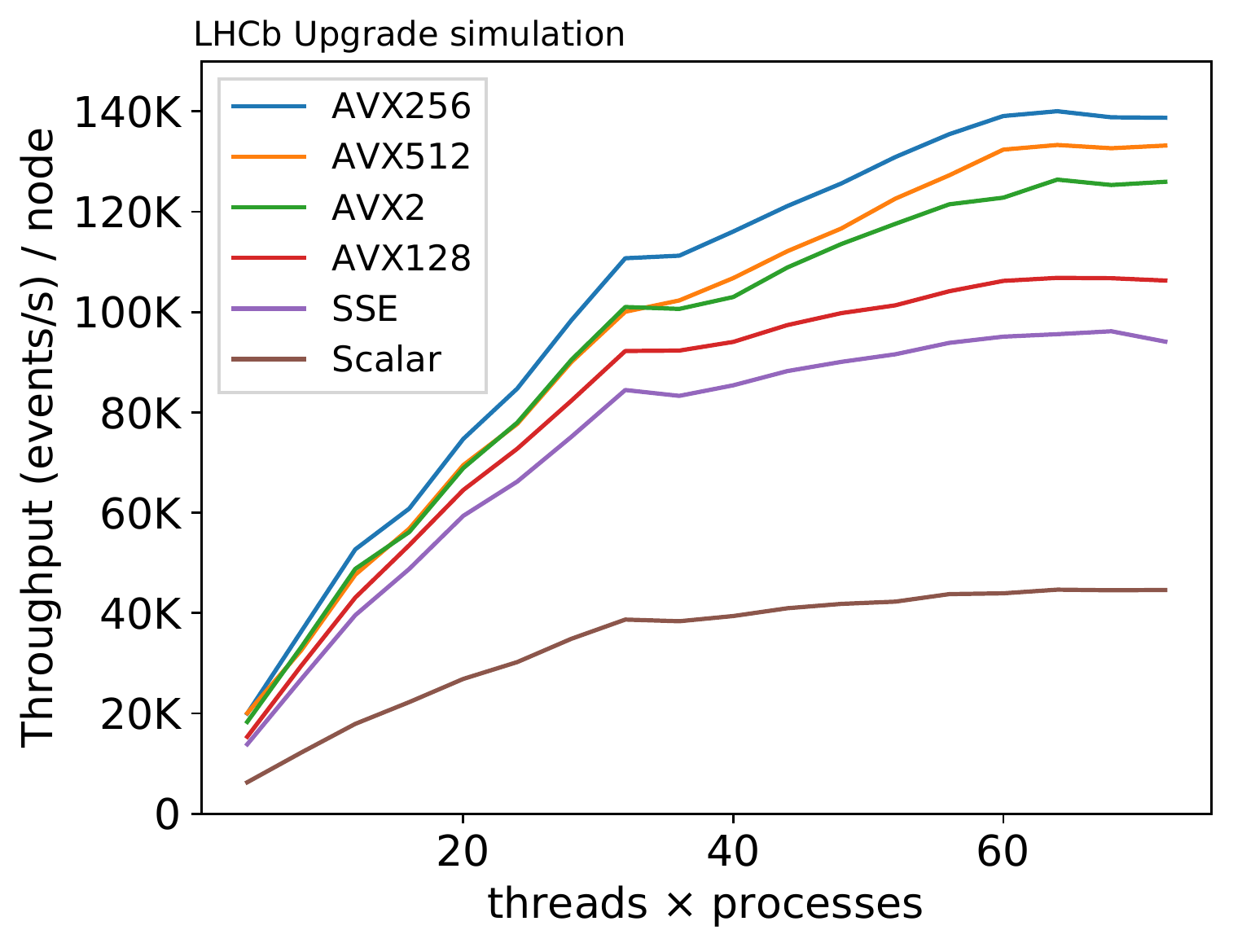}
\caption{Throughput as a function of the number of threads for two processes (one on each NUMA domain) on dual-socket Intel Xeon Gold 6130. On the left: comparison of SIMD VELO Tracking with the Search by Pair (SbP) algorithm in "fast" and "best" configurations. On the right: comparison of different SIMD backends.}
\label{fig:scaling_threads}
\end{center}
\end{figure}

\begin{figure}[h!]
\begin{center}
\includegraphics[scale=0.47]{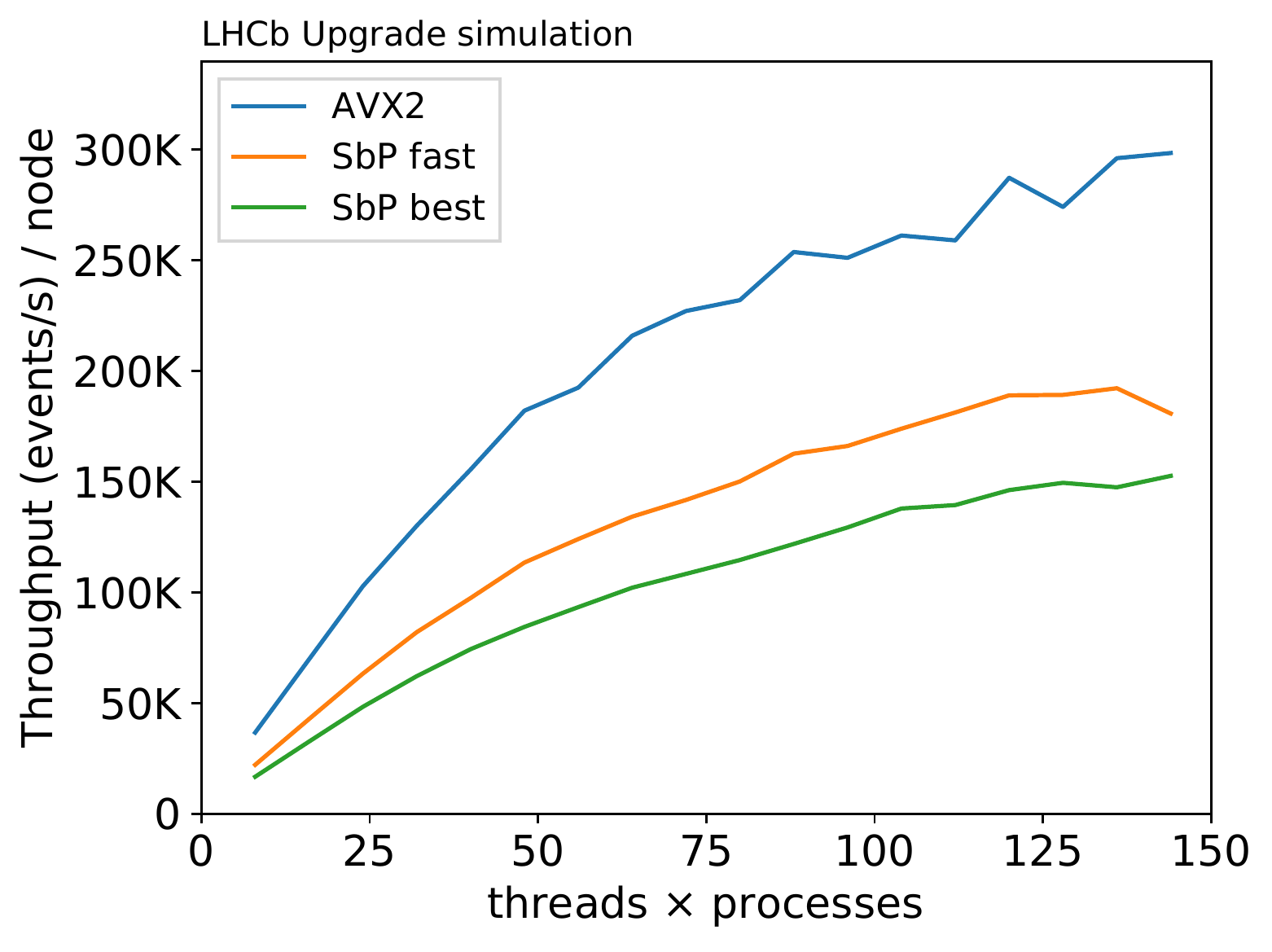}
\includegraphics[scale=0.47]{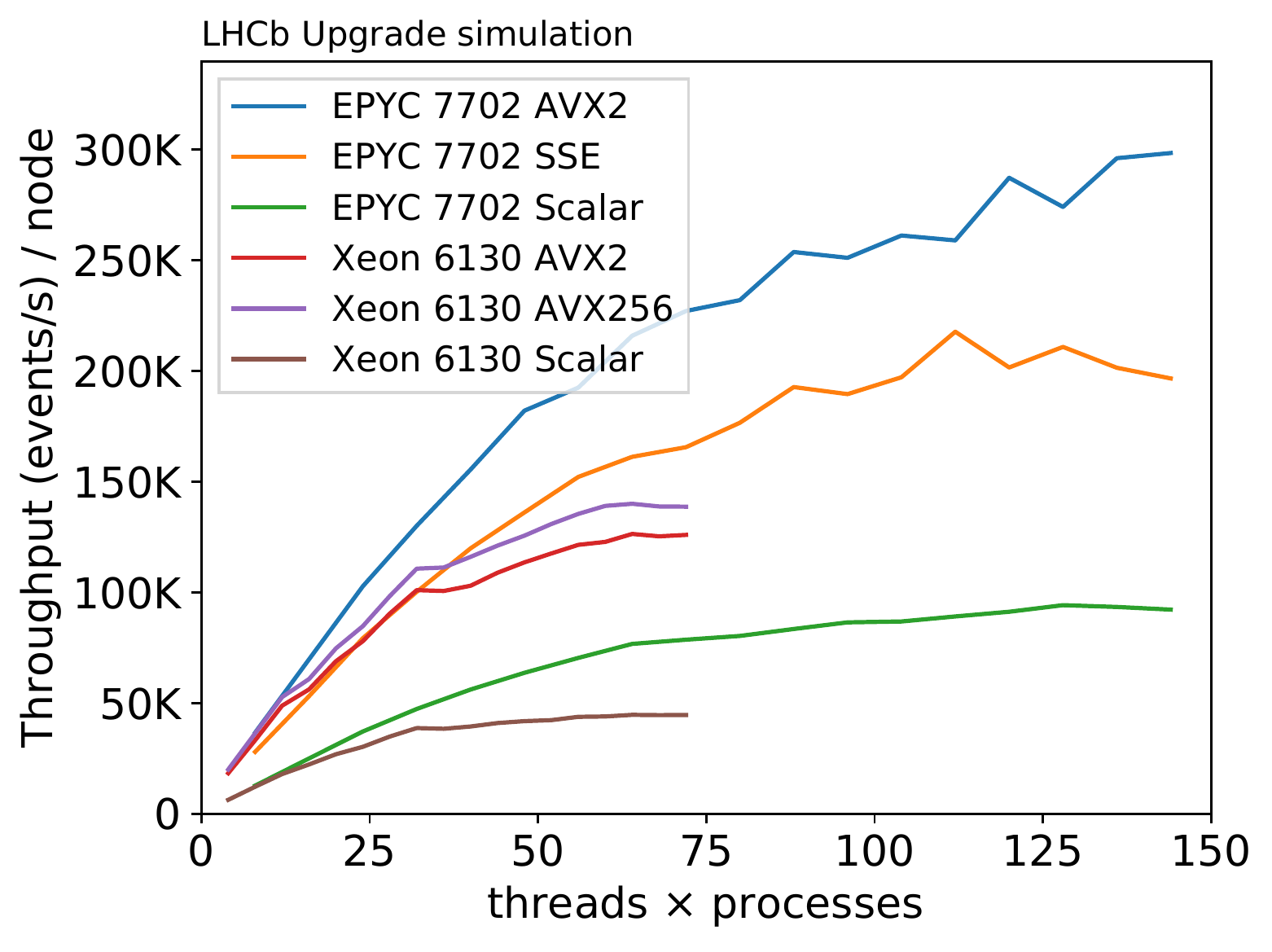}
\caption{On the left: comparison of SIMD VELO Tracking with the Search by Pair (SbP) algorithm in "fast" and "best" configurations, on a single socket AMD EPYC "Rome" 7702. On the right: comparison of a single socket AMD EPYC "Rome" 7702 and a dual-socket Intel Xeon Gold 6130 for different SIMD backends.}
\label{fig:scaling_amd_threads}
\end{center}
\end{figure}

All throughput tests were done on minimum bias Monte-Carlo simulation samples. First, the throughput of the new SIMD VELO Tracking algorithm is compared with the 2018 Search by Pair (SbP) algorithm, the current state-of-the-art for VELO pattern reconstruction on CPU. The SbP algorithm was originaly coming in two versions: the "fast" configuration favoring speed over efficiency was meant to be used for HLT1 and the "best" configuration favoring efficiency over speed was meant to be used for HLT2. The left side of Figures~\ref{fig:scaling_threads} and \ref{fig:scaling_amd_threads} shows the SIMD algorithm is \emph{faster} than both the "fast" and "best" configurations of SbP, on \emph{every} tested architectures, for any number of threads. Using SIMD wrappers, different implementations of the SIMD algorithm were also compared. Interestingly, the AVX512 backend with an SIMD register width of 16 performs less well than the AVX256 with a register width of 8. This can be explained by the frequency scaling issues discussed in Section~\ref{sec:SIMD}. However, thanks to the new instructions introduced with AVX-512, the AVX256 and AVX128 backends bring a 10\% improvement over plain AVX2 and SSE respectively. The scalar backend is significantly lower than all other SIMD backends because the compiler is not able to vectorize the filtering pattern. The slow down in speedup progression with increased SIMD parallelism is a combination of frequency scaling and not being able to extract enough parallelism from relatively small loops. The right of Figure~\ref{fig:scaling_amd_threads} presents a comparison of the Intel and AMD systems for the relevant backends. Because the Intel setup only have 32 physical cores, the two architecture can only be compared at threads $\times$ processes $=32$. The AMD's AVX2 backend provides a 28\% improvement over Intel's AVX2 and a 18\% improvement over Intel's AVX256, despite the frequency being 20\% lower for AMD. AMD's scalar backend also increased the throughput by 23\% from Intel's scalar backend.

\begin{figure}[h!]
\centering
\includegraphics[scale=0.47]{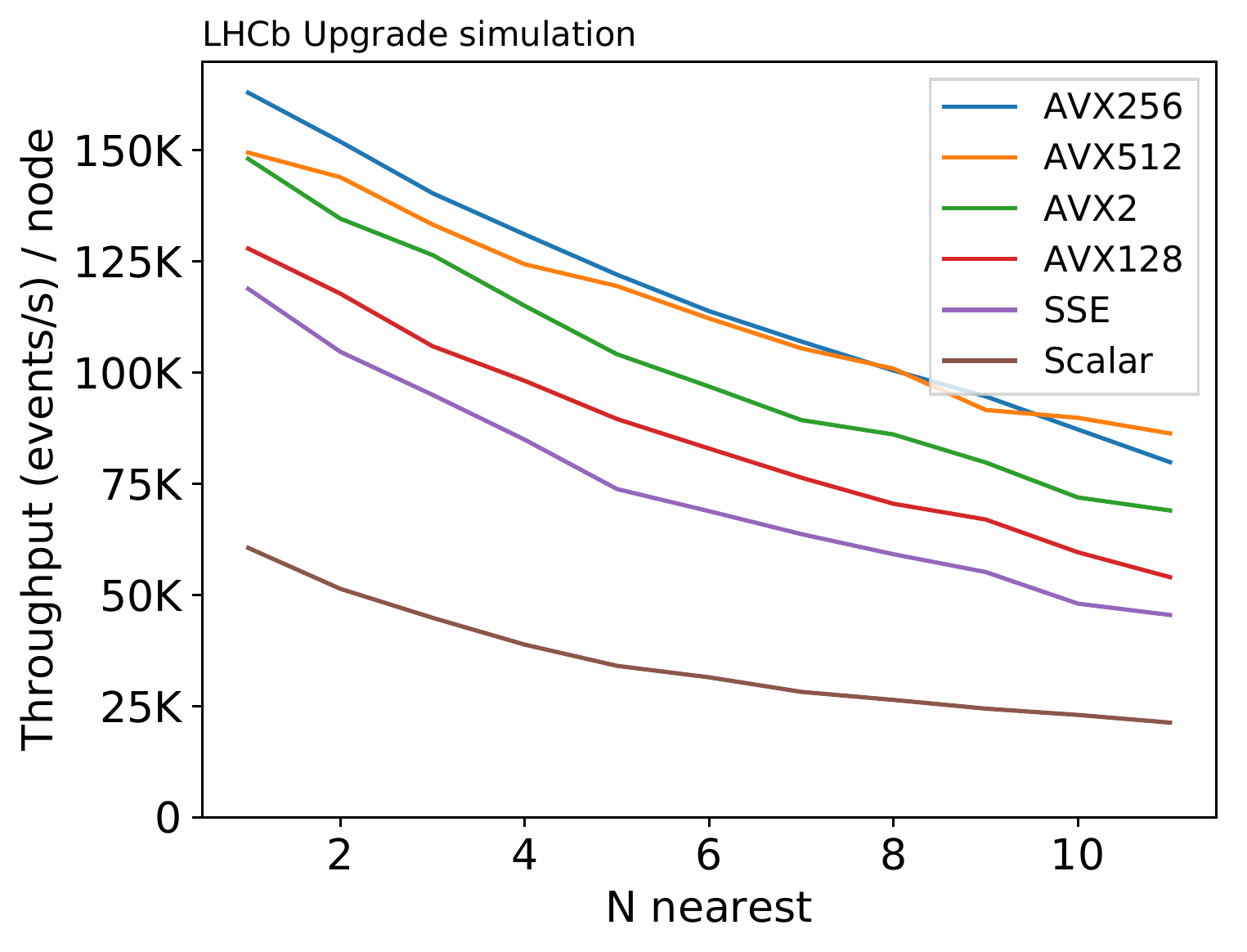}
\includegraphics[scale=0.47]{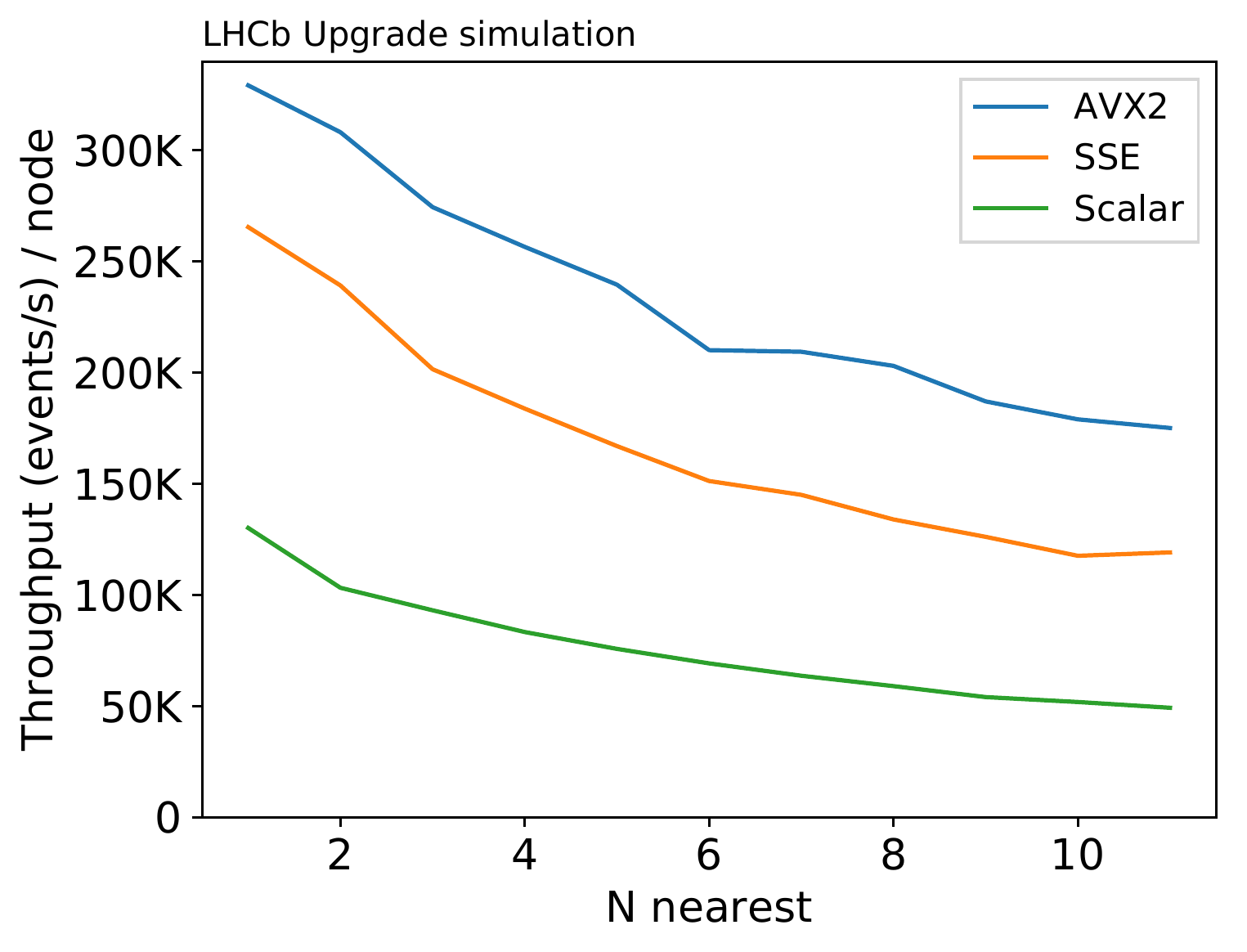}
\caption{Throughput as a function of the number of pair candidates in the track seeding step, for different SIMD backends. On the left: dual-socket Intel Xeon Gold 6130. On the right: single socket AMD EPYC "Rome" 7702.}
\label{fig:scaling_N}
\end{figure}

As the throughput and efficiency of the algorithm depends linearly from the number of pair candidates to consider during the seeding step, it offers a direct tuning parameter to adjust the trade-off between speed and efficiency. Figure~\ref{fig:scaling_N} shows the progression of throughput as a function of the number of pair candidates.

\subsection{Reconstruction efficiency}

While minimum bias events are representative of the data seen by the production system in real-time, the majority of them do not contain the physical signals whose efficiency we want to optimize. We therefore simulate a typical signal of interest for LHCb to measure the algorithm efficiencies. We choose $B_s\rightarrow\phi\phi$, which is the same signal used in all LHCb historical publications on tracking. 

\begin{figure}[h]
\centering
\includegraphics[scale=0.5]{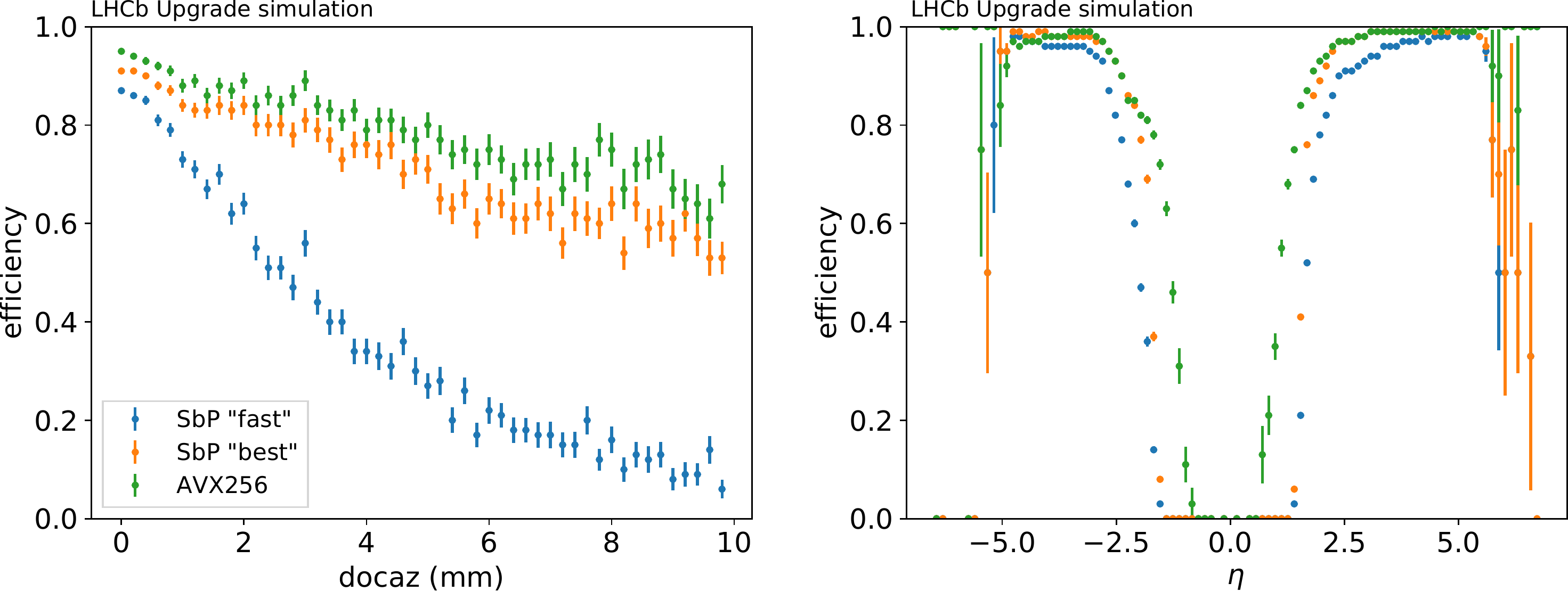}
\caption{Efficiency as a function of the distance of closest approach to the z axis (docaz), in mm, and the pseudorapidity $\eta$ for Velo tracks. LHCb is mostly interested in tracks with DOCAZ < 1 mm.}
\label{fig:eff_docaz_eta}
\end{figure}

We present the efficiency of different track categories as a function of multiple physical parameters of interest. The efficiency as a function of the distance of closest approach to beamline (docaz) is studied to ensure the efficient reconstruction of tracks produced in the decays of long-lived particles. Because of the geometry of the VELO, it is also particularly interesting to plot the efficiency as a function of track pseudorapidity ($\eta$).\footnote{The pseudorapidity of a track is given by $-\ln\left[\tan\left(\frac{\theta}{2}\right)\right]$, where $\theta$ is the angle of the track to the beamline} Figure~\ref{fig:eff_docaz_eta} shows the efficiency as a function of docaz and $\eta$ for the two configurations of the Search by Pair algorithm and the AVX256 version of the SIMD VELO Tracking. The docaz efficiencies are plotted in the range $2<\eta<5$, which represents the acceptance of the full LHCb detector. Thanks to its nearest $\phi$ approach, the SIMD VELO Tracking is more efficient than previous State-of-the-Art for very displaced tracks, even if the number of evaluated pair candidates is small (N=3). It also has significantly better efficiencies for very small track $\eta$. While small $\eta$ tracks do not pass through the rest of the detector, they can nevertheless play an important role in the reconstruction of PVs, but a detailed study of this is left to a future publication.

We also evaluate the integrated efficiency in the range $2<\eta<5$ for different track types. We distinguish between:
\begin{itemize}
    \item all tracks that leaves at least 3 hits in the VELO detector ("VELO tracks") which are mostly produced directly in the PVs,
    \item tracks that come from the decay of a hadron containing a bottom quark and traverse the rest of the LHCb tracking detectors ("From B"),
    \item tracks that come from the decay of a hadron containing a charm quark and traverse the rest of the LHCb tracking detectors ("From D"),
    \item tracks that come from the decay of a hadron containing a strange quark and traverse the rest of the LHCb tracking detectors ("Strange").
\end{itemize}
Table~\ref{table:eff} compares the efficiencies for these categories and compares the fake rate for the two configurations of SbP and AVX256 SIMD VELO Tracking. More efficient algorithms tends to produce more fakes due to the higher number of combinations tested. Still, the SIMD algorithm produces fewer fakes than the "best" SbP. The SIMD algorithm is more efficient for "VELO" and "From B" categories, while being within 1\% of the "best" SbP for "From D" and "Strange" categories. While the results presented here only allow for $N=3$ pair candidate, the algorithm can outperform the other algorithms in all categories for $N\geq6$.  The clone rates are similarly presented in Table~\ref{table:clones}, and the SIMD algorithm produces fewer clones than previous approaches.

\begin{table}[h]
\centering
\begin{tabular}{l|l|llll}
\hline
                              & Fakes        & Velo           & From B          & From D          & Strange \\ \hline
Search by Pair "fast" & \textbf{0.83}          & 93.05          & 95.64          & 95.29          & 79.34 \\
Search by Pair "best"    & 1.22          & 97.62          & 98.71          & \textbf{99.05} & \textbf{97.46} \\
VELO Tracking SIMD       & 1.04          & \textbf{98.20} & \textbf{99.12} & 98.99          & 96.82 \\ \hline
\end{tabular}
\caption{Efficiencies for tracks that are not electrons in the range $2<\eta<5$.}
\label{table:eff}
\end{table}

\begin{table}[h]
\centering
\begin{tabular}{l|llll}
\hline
                              & \begin{tabular}[c]{@{}l@{}}Velo\\ clones\end{tabular} & \begin{tabular}[c]{@{}l@{}}From B\\ clones\end{tabular} & \begin{tabular}[c]{@{}l@{}}From D\\ clones\end{tabular} & \begin{tabular}[c]{@{}l@{}}Strange\\ clones\end{tabular} \\ \hline
Search by Pair "fast" & 2.31                                                  & 0.89                                                   & 1.42                                                   & 1.54                                                     \\
Search by Pair "best"    & 2.75                                                  & 0.84                                                   & 1.25                                                   & \textbf{0.82}                                            \\
VELO Tracking SIMD       & \textbf{1.35}                                         & \textbf{0.68}                                          & \textbf{0.90}                                          & \textbf{0.82}                                            \\ \hline
\end{tabular}
\caption{Clone rates (in \%) on 1000 $B_S\rightarrow\phi\phi$ events, for $2<\eta<5$ tracks. The lower, the better.}
\label{table:clones}
\end{table}

\begin{figure}[h]
\centering
\includegraphics[scale=0.48]{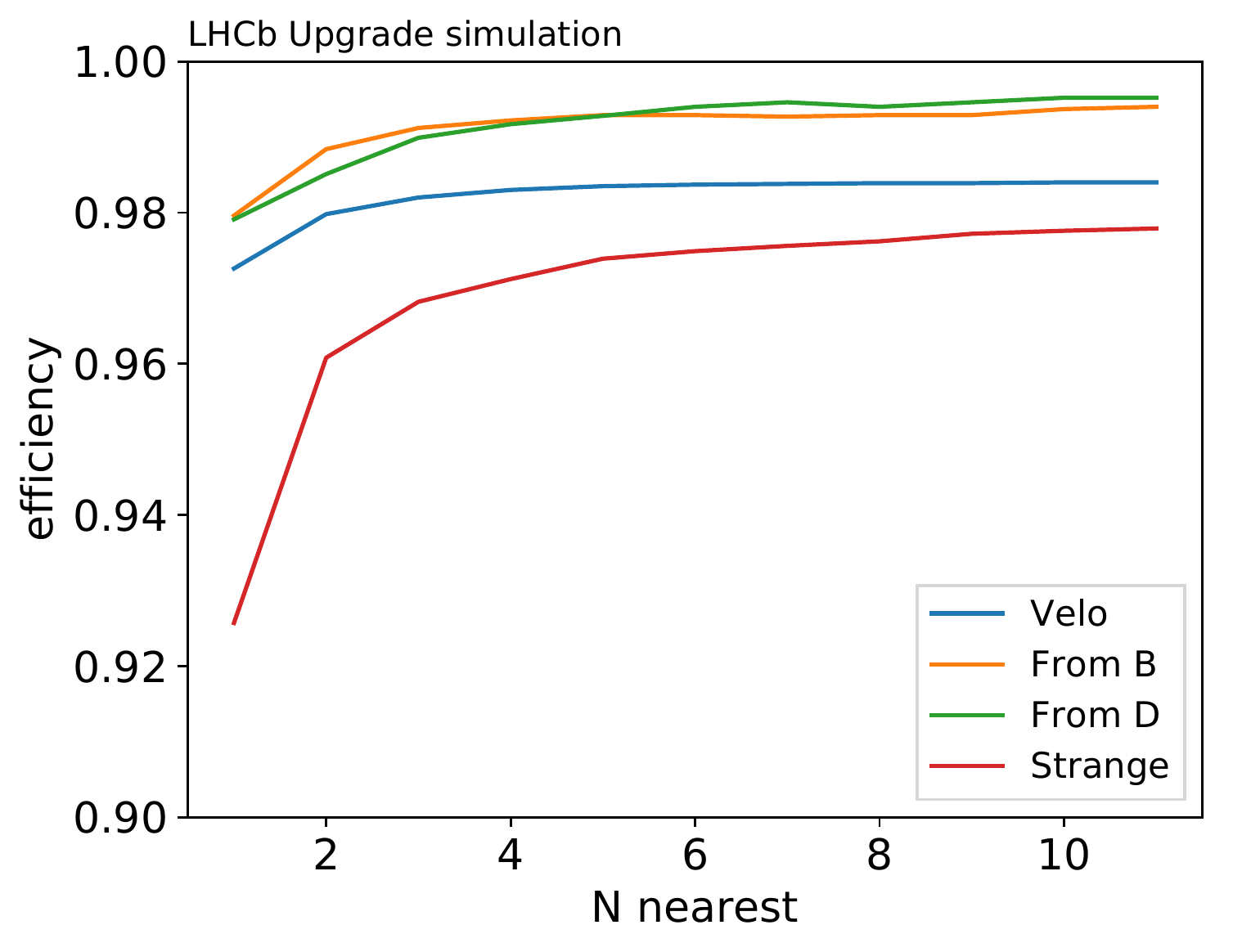}
\includegraphics[scale=0.48]{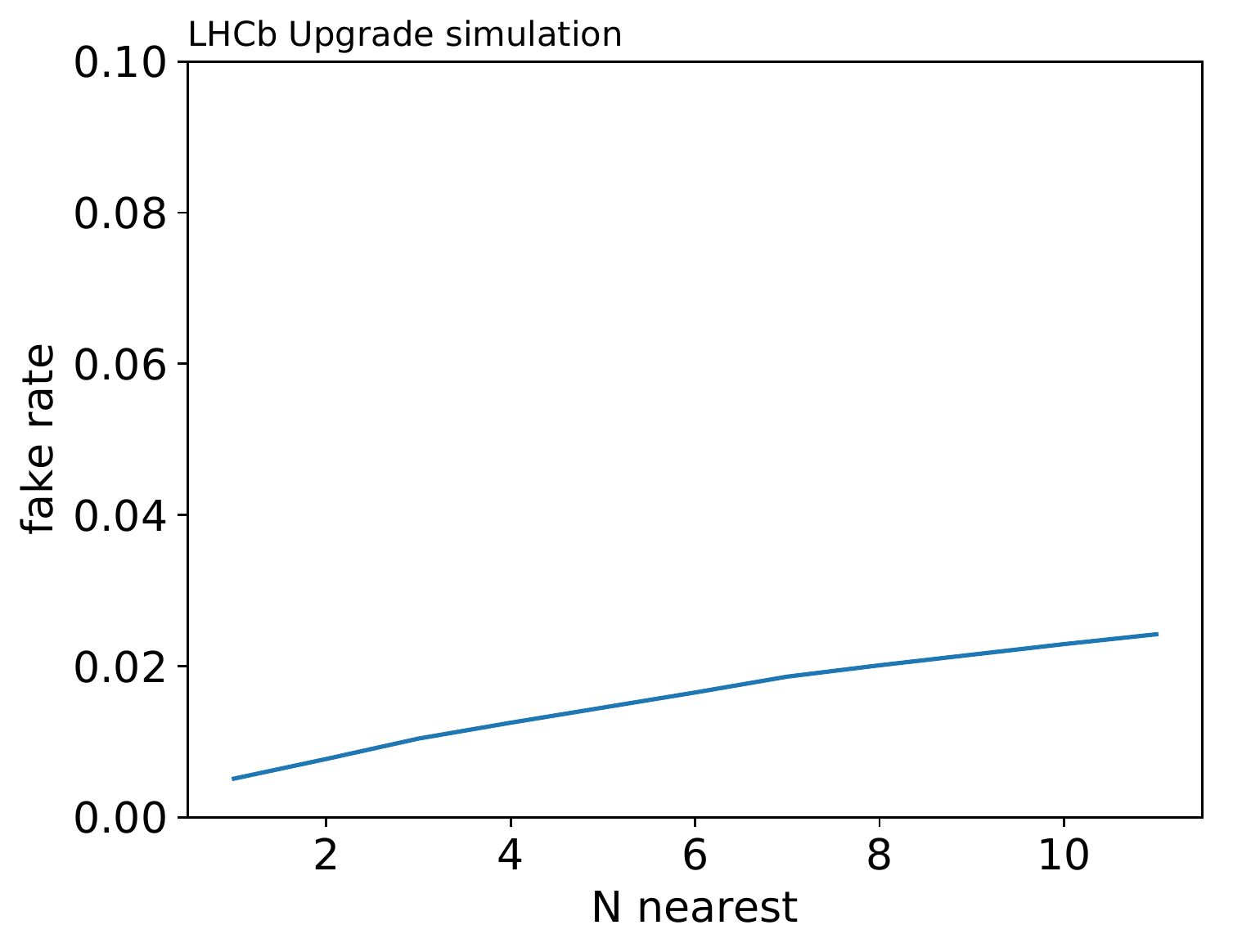}
\caption{Efficiencies for the different track categories described in the text as a function of the number of seeding candidates. Higher efficiencies and lower fake rate are better.}
\label{fig:eff_N}
\end{figure}

Figure~\ref{fig:eff_N} shows the impact of varying the number of seeding pair candidates on the efficiency and fake rate. As N increases, the efficiencies go up, but as more pairs are tested, the probability of finding randomly aligned hits also increases leading to more fakes. We noted that already at N=1, the efficiencies are higher than the "fast" SbP algorithm and could offer a viable mitigation for HLT1. However, to minimize the disparity between HLT1 and HLT2 track reconstructions, we choose a default value of N=3 for both configurations as it seems to be a good trade-off between throughput, efficiency and fake rate.\\

Table~\ref{table:hit_eff} shows the hit efficiencies for the same track categories. The "All hits efficiencies" section shows that in average, the SIMD algorithm finds more correct hits than the SbP algorithms. The "First 3 hits" and "last hit" categories are important to correctly extrapolate the track to the beamline or to the end of the VELO detector. Again, the SIMD algorithm demonstrate better, or within 1\%, efficiencies than the SbP algorithms.\\

\begin{table}[h]
\centering
\begin{tabular}{l|llll}
\hline
                              & Velo           & From B          & From D          & Strange \\ \hline
\multicolumn{5}{l}{All hits efficiencies:}\\
\hline
Search by Pair "fast" & 90.05          & 92.40          & 92.82          & 84.69 \\
Search by Pair "best" & 94.19          & 97.61          & 97.50          & 97.47 \\
VELO Tracking SIMD    & \textbf{96.97} & \textbf{97.96} & \textbf{98.05}          & \textbf{97.58} \\
\hline
\multicolumn{5}{l}{Efficiencies of first 3 hits:}\\
\hline
Search by Pair "fast" & 91.74          & 93.93          & 94.32          & 76.54 \\
Search by Pair "best" & 93.86          & 97.44          & 97.54          & 96.80 \\
VELO Tracking SIMD    & \textbf{97.08} & \textbf{98.18} & \textbf{98.13}          & \textbf{97.39} \\
\hline
\multicolumn{5}{l}{Efficiencies of last hit:}\\
\hline
Search by Pair "fast" & 87.95          & 90.77          & 91.43          & 88.97 \\
Search by Pair "best" & 94.07          & \textbf{97.64}          & 97.48          & \textbf{97.72} \\
VELO Tracking SIMD    & \textbf{96.60} & 97.53 & \textbf{97.85}          & 97.12 \\ \hline
\end{tabular}
\caption{Average hit efficiencies of reconstructed tracks in the range $2<\eta<5$. All results are given for an SIMD register width of 8. (Best efficiencies in bold)}
\label{table:hit_eff}
\end{table}

\section{Conclusion}

In this article, we have presented a new tracking algorithm for the VELO detector of the LHCb experiment specialized to take advantage of SIMD general purpose multicore processors. We compared it to previous State-of-the-Art pattern recognition algorithms and showed a significant speedup and in some cases increase in efficiency over all previous alternatives. This allows the SIMD algorithm to be used for all stages of LHCb's real-time data processing. We also evaluated the algorithm on two high-end systems from Intel and AMD and showed the impact of the SIMD extensions on the performance.


\acknowledgments

The authors would like to thank the Physics Data Processing group from Nikhef and in particular Tristan Suerink, for lending them the AMD EPYC "Rome" 7702 system and Gvozden Ne\v{s}kovi\'{c} from the Frankfurt Institute for Advanced Studies for the opportunity to have an early look at the AMD EPYC "Rome" architecture. They would also like to thank the LHCb computing and simulation teams for their support and for producing the simulated LHCb samples used in the paper. VVG, RQ, and AH are partially supported by ERC-CoG-724777 ``RECEPT''.




\bibliographystyle{ieeetr}

\end{document}